\documentclass[prb,twocolumn,superscriptaddress,showpacs,nofootinbib,notitlepage]{revtex4-1}
\usepackage[utf8]{inputenc}

\usepackage{graphicx,color,dsfont,xcolor,amsmath, amssymb}
\usepackage{gensymb}
\usepackage{hyperref}
\usepackage[caption=false]{subfig} 
\usepackage[normalem]{ulem}
\usepackage[titletoc,toc,title]{appendix}
\usepackage{bbold}

\DeclareMathOperator{\im}{im}

\definecolor{darkblue}{RGB}{0,0,127}

\newcommand{\ket}[1]{|#1\rangle}
    
\usepackage[all]{xy}
\newcommand{\drawgenerator}[8]{%
\xymatrix@!0{%
& #8 \ar@{-}[ld]\ar@{.}[dd] \ar@{-}[rr] & & #7 \ar@{-}[ld]  \\%
#1 \ar@{-}[rr] \ar@{-}[dd] &  & #2 \ar@{-}[dd] &            \\%
& #6 \ar@{.}[ld] &  & #5 \ar@{-}[uu] \ar@{.}[ll]       \\%
#3 \ar@{-}[rr] &  & #4 \ar@{-}[ru]                       %
}%
}
\newcommand{\drawsolidgenerator}[8]{%
\xymatrix@!0{%
& #8 \ar@{-}[ld]\ar@{-}[d] \ar@{-}[rr] & & #7 \ar@{-}[ld]  \\%
#1 \ar@{-}[rr] \ar@{-}[dd] & \ar@{-}[d] & #2 \ar@{-}[dd] &            \\%
& #6 \ar@{-}[ld] & \ar@{-}[l]   & #5 \ar@{-}[uu] \ar@{-}[l]       \\%
#3 \ar@{-}[rr] &  & #4 \ar@{-}[ru]                       %
}%
}
\newcommand{\drawbig}[9]{%
\xymatrix@!0@=.66cm{%
& #8 \ar@{-}[ld]\ar@{-}[d] \ar@{-}[rrr] & & & #7 \ar@{-}[ld]  \\%
#1 \ar@{-}[rrr] \ar@{-}[ddd] & & & #2 \ar@{-}[ddd] &            \\%
&  & #9 &  &       \\%
& #6 \ar@{-}[ld] \ar@{-}[rr] \ar@{-}[uu]  & &   & #5 \ar@{-}[uuu] \ar@{-}[l]       \\%
#3 \ar@{-}[rrr] & & & #4 \ar@{-}[ru]                       %
}%
}
\newcommand{\drawbigdual}[9]{%
\xymatrix@!0@=.66cm{%
& #8 \ar@{.}[ld]\ar@{.}[d] \ar@{.}[rrr] & & & #7 \ar@{.}[ld]  \\%
#1 \ar@{.}[rrr] \ar@{.}[ddd] & &  & #2 \ar@{.}[ddd] &            \\%
&  & #9  &  &     \\%
& #6 \ar@{.}[ld] \ar@{.}[rr] \ar@{.}[uu]  &   &   & #5 \ar@{.}[uuu] \ar@{.}[l]       \\%
#3 \ar@{.}[rrr] & & & #4 \ar@{.}[ru]                       %
}%
}

\hypersetup{
  colorlinks = true,
  urlcolor = blue,
  pdfauthor = {Dominic J. Williamson and Trithep Devakul},
  pdftitle = {Type-II fractons from coupled cluster state chains and layers}
}

\begin{document}

\title{Type-II fractons from coupled spin chains and layers 
}

\date{\today}

\author{Dominic J. Williamson}
\affiliation{Stanford Institute for Theoretical Physics, Stanford University, Stanford, CA 94305, USA}
\author{Trithep Devakul}
\affiliation{Department of Physics, Princeton University, NJ 08540, USA}

\begin{abstract} 
We describe a construction of topological orders from coupled lower dimensional symmetry-protected topological orders, which is closely related to gauging a subsystem symmetry. 
Our construction yields both conventional topological orders and exotic fracton topological orders of type-I and type-II. 
In particular, we find a coupled spin chain construction of Haah's cubic code, and a coupled layer construction of Yoshida's fractal spin liquids. 
\end{abstract}

\maketitle

Fracton topological order is an exotic type of quantum order characterized by topological charges with restricted mobility due to superselection rules. 
To date, a wide range of 3D exactly solved models exhibiting fracton order have been discovered~\cite{chamon2005quantum,haah2011local,doi:10.1080/14786435.2011.609152,kim20123d,yoshida2013exotic,haah2014bifurcation,PhysRevB.92.235136,vijay2016fracton,vijay2017generalization,prem2018cage,song2018twisted,Prem2019,Bulmash2019,Tantivasadakarn2019,Devakul2020,Stephen2020,Tantivasadakarn2020,Shirley2020,Aasen2020,Williamson2020,Fontana2020}. 
The limited mobility of topological charges in these models has lead to fracton topological codes with unique advantages over conventional topological codes, whose charges are fully mobile~\cite{PhysRevLett.107.150504,bravyi2013quantum,Brown2019}. 
First and foremost are type-II fracton codes, which support only immobile topological charges, as opposed to type-I fracton codes, which support charges with less severe mobility restrictions. 
It has been shown that the slow thermal dynamics of topological charges in Haah's cubic code, the canonical type-II model, lead to a partially self-correcting quantum memory~\cite{bravyi2013quantum}. 

The implementation of quantum error correction is a vital step that must be overcome to engineer scalable quantum computers. 
The leading approaches to quantum error correction are dominated by conventional topological codes. Looking forward, fracton topological codes provide a bevy of options, some promising better scaling of encoded qubits, code distance and memory time with system size. 
However, even the simplest class of fracton models, those described by commuting Pauli stabilizer Hamiltonians\footnote{see the appendix of Ref.~\onlinecite{Dua2019a}  for a repository of models}, require high degree interactions. This poses a significant barrier to any potential experimental realizations of fracton phases. 
To address this issue, several constructions of fracton models from coupled lower dimensional layers~\cite{PhysRevB.95.245126,vijay2017isotropic,PhysRevB.96.165106,Fuji2019,Schmitz2019,Shirley2019,Aasen2020,Williamson2020,Wen2020,Wang2020} 
or spin chains~\cite{PhysRevLett.119.257202,PhysRevB.96.165105} have been found, some with much lower degree interactions. 

In this work we introduce a construction of fracton stabilizer models from coupled cluster state~\cite{cluster} layers, and spin chains, which are symmetry-protected topological (SPT) orders~\cite{chen2012symmetry}. 
In particular, we describe a construction of Haah's cubic code~\cite{Haah}
from coupled spin chains, featuring at most four body interactions. 
To the best of our knowledge\footnote{We were not able to find a proof that the ``type-II'' model in Ref.~\onlinecite{PhysRevB.96.165105} supports no nontrivial string operators~\cite{haah2011local} and therefore satisfies the definition of type-II~\cite{vijay2016fracton}.} 
this is the first coupled spin chain construction of a model that has been rigorously shown to be type-II. 
We go on to explain how our coupled cluster state construction is related to the known construction of fracton models from gauging subsystem symmetries~\cite{vijay2016fracton,PhysRevB,shirley2018FoliatedFracton,kubica2018ungauging}. 
Our construction can be implemented by entangling spin systems, arranged in chains or layers, with auxiliary quantum degrees of freedom that are then destructively measured. 

We open in section~\ref{sec:examples} with several examples obtained via the general method explained in section~\ref{sec:gauging} and~\ref{sec:CoupledLayers}, then discussing an implementation of the construction in a modular architecture in section~\ref{sec:modular}. 
Further examples are presented in appendix~\ref{app:examples}.

\section{Examples} 
\label{sec:examples}
Here we lead with examples, as they can be presented before explaining the general construction. 
First we describe a coupled cluster chain construction of the 2D toric code~\cite{kitaev}  and Haah's cubic code. We then describe a coupled cluster layer construction of Yoshida's type-II $\mathbb{Z}_3$ Fractal spin liquid (FSL), which generalizes to all FSL models~\cite{yoshida2013exotic} (see the appendix for this and further examples). 
In the sections that follow we explain how these examples were found as instances of our general construction. 

The examples in this paper make use of a description of Pauli stabilizer Hamiltonians in terms of a set of local generators $\{h_i\}_i$ for the stabilizer group $\langle \{h_i\} \rangle $. The commuting projector Hamiltonian whose ground space is the stabilizer codespace is given by the sum of generators 
\begin{align}
    H = - \sum_i h_i
    \, .
\end{align}

The 2D toric code Hamiltonian~\cite{kitaev}, governing qubits on the edges of a square lattice, is described by translates of the local stabilizer generators 
\begin{align}
\begin{array}{c}
\xymatrix@!0@=.8cm{%
& Z 
\\
Z &  \ar@{-}[u]  \ar@{-}[d]  \ar@{-}[l]  \ar@{-}[r]  & Z
\\
& Z
}
\end{array},
&&
\begin{array}{c}
\xymatrix@!0@=.8cm{%
  & X \ar@{-}[l]  \ar@{-}[r]  & 
\\
 X  \ar@{-}[u]  \ar@{-}[d]   & & X \ar@{-}[u]  \ar@{-}[d]
\\
  & X \ar@{-}[l]  \ar@{-}[r] & 
}
\end{array},
\label{eq:2DTC}
\end{align}
where $X$, $Z$ are Pauli matrices.
This Hamiltonian can be constructed by coupling 1D cluster states~\cite{cluster} along the horizontal and vertical directions. 
The stabilizer generators for these cluster states are given by translates of 
\begin{align}
\begin{array}{c}
\xymatrix@!0@=.8cm{%
& 
\\
Z   & XI  \ar@{-}[u]  \ar@{-}[d]  \ar@{-}[l]  \ar@{-}[r]    & Z
\\
&
}
\end{array},
\begin{array}{c}
\xymatrix@!0@=.8cm{%
ZI   & X  \ar@{-}[l] \ar@{-}[r]  & ZI
}
\end{array},
\begin{array}{c}
\xymatrix@!0@=.8cm{%
& Z 
\\
& IX \ar@{-}[u]  \ar@{-}[d]  \ar@{-}[l]  \ar@{-}[r]  & 
\\
& Z
}
\end{array},
\begin{array}{c}
\xymatrix@!0@=.8cm{%
 IZ 
\\
 X  \ar@{-}[u]  \ar@{-}[d] 
\\
 IZ  
 \\
}
\end{array},
\nonumber
\end{align}
where each vertex hosts two qubits, and each edge hosts a single qubit. 
We have utilized compact notation for operators on the multi-qubit vertices, e.g. $XZ$ denotes $X \otimes Z$. 
The first and third terms in the above equation act on vertex qubits and a pair of adjacent edge qubits, while the second and fourth terms act on an horizontal or vertical edge qubit and a pair of adjacent vertex qubits. 
In the limit of a strong $-ZZ$ field on all vertices we recover the vertex-edge cluster state on the square lattice. 
Similarly in the limit of a strong $-(XX+ZZ)$ field on all vertices we find the 2D toric code Hamiltonian in Eq.~\eqref{eq:2DTC} at leading order in perturbation theory.

Haah's cubic code is described by translates of the following local Hamiltonian terms, or stabilizer generators, 
\begin{align}
\begin{array}{c}
\drawsolidgenerator{ZI}{ZZ}{IZ}{ZI}{IZ}{II}{ZI}{IZ}
\end{array},
&&
\begin{array}{c}
\drawsolidgenerator{XI}{II}{IX}{XI}{IX}{XX}{XI}{IX}
\end{array} .
\label{eq:CC}
\end{align}
Surprisingly, we have found that this model can be obtained by coupling cluster chains stacked along three different directions. The Hamiltonians for these cluster chains are given by translates of the following local terms\begin{align}
\begin{array}{c}
\drawbig{{}}{{ZIII}}{{}}{{ZIII}}{{}}{{}}{{}}{{}}{{XIII}}
\end{array},
&&
\begin{array}{c}
\drawbigdual{{}}{{}}{{}}{}{{}}{{ZIII}}{{}}{{ZIII}}{{XIII}} 
\end{array},
\nonumber
\\
\begin{array}{c}
\drawbig{{IZII}}{{}}{{}}{{}}{{}}{{}}{{IZII}}{{}}{{IXII}}
\end{array},
&&
\begin{array}{c}
\drawbigdual{{}}{{}}{{IZII}}{{}}{{IZII}}{{}}{{}}{{}}{{IXII}} 
\end{array},
\nonumber
\\
\begin{array}{c}
\drawbig{{}}{{IIZI}}{{}}{{}}{{}}{{}}{{}}{{IIZI}}{{IIXI}}
\end{array},
&&
\begin{array}{c}
\drawbigdual{{}}{{}}{{}}{{IIZI}}{{}}{{IIZI}}{{}}{{}}{{IIXI}}
\end{array},
\nonumber
\\
\begin{array}{c}
\drawbig{{}}{{}}{{IIIZ}}{{}}{{IIIZ}}{{}}{{}}{{}}{{IIIX}}
\end{array},
&&
\begin{array}{c}
\drawbigdual{{IIIZ}}{{}}{{}}{{}}{{}}{{}}{{IIIZ}}{{}}{{IIIX}}
\end{array},
\nonumber
\end{align}
where each vertex and cube center now hosts four qubits. The dotted lines indicate the dual lattice, whose vertices lie on the cube centers of the original lattice (solid lines), and vice versa. Each pair of terms on a line in the equation above generates a cluster chain that zigzags between sites of the lattice and dual lattice along the $\hat{y}$ or $\hat{z}\pm\hat{x}$ direction. 
In the limit of a strong $-(ZZII+IIZZ)$ field on the vertices and $-(ZZII+IZZI+IIZZ+XXXX)$ on the cube centers (dual vertices) we recover the cubic code Hamiltonian in Eq.~\eqref{eq:CC} at leading order in perturbation theory.

\begin{figure}[t!]
    \centering
    \subfloat[\label{fig:xzLayers}]{\includegraphics[width=0.3\columnwidth]{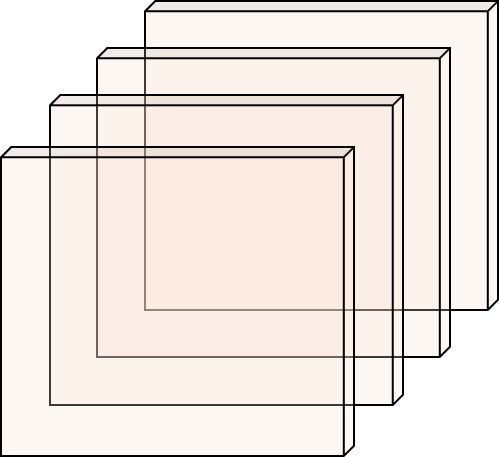}}
    \hspace{.1cm}
    \subfloat[\label{fig:xyLayers}]{\includegraphics[width=0.3\columnwidth]{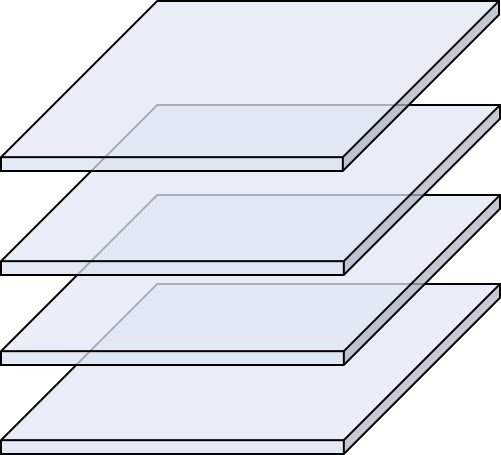}}
    \hspace{.1cm}
    \subfloat[\label{fig:xyzLayers}]{\includegraphics[width=0.3\columnwidth]{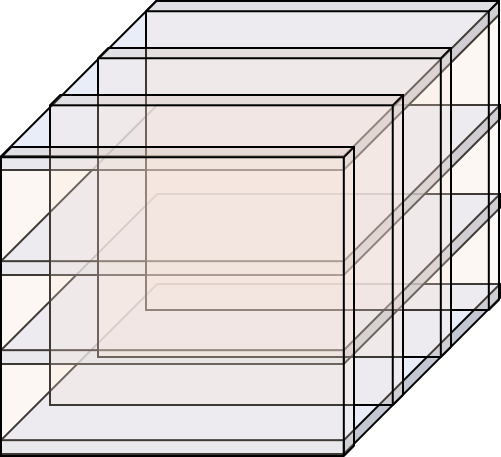}}
    \caption{
    (a) A stack of fractal cluster states in $xy$ layers. 
    (b) A stack of fractal cluster states in $xz$ layers. 
    (c) An FSL model obtained by coupling the $xy$ and $xz$ layers. 
}
\label{fig:FSL}
\end{figure}

The Hamiltonian for Yoshida's type-II $\mathbb{Z}_3$ FSL model is a sum of translates of the following local generators
\begin{align}
&
\begin{array}{c}
\xymatrix@!0@=.66cm{%
&  \ar@{-}[ld]\ar@{-}[d] \ar@{-}[rr] & &  \ar@{-}[ld] & &  \ar@{-}[ld]  \ar@{-}[ll]  \ar@{-}[dd]  
\\%
IZ \ar@{-}[rr] \ar@{-}[dd] & \ar@{-}[d]   &  \ar@{-}[dd]  \ar@{-}[rr] &  \ar@{-}[u] & IZ  &        
\\%
 & ZI  \ar@{-}[ld] & \ar@{-}[l] &  ZI \ar@{-}[u] \ar@{-}[l]  \ar@{-}[r]  & \ar@{-}[r] &    
 \\%
ZZ \ar@{-}[rr] &  &  \ar@{-}[ru]  & &  \ar@{-}[ll]  \ar@{-}[uu]  \ar@{-}[ur] %
}%
\end{array},
\begin{array}{c}
\xymatrix@!0@=.66cm{%
&  \ar@{-}[ld]\ar@{-}[d] \ar@{-}[rr] & & \ar@{-}[ld] & & XX^\dagger \ar@{-}[ld]  \ar@{-}[ll]  \ar@{-}[dd]  
\\%
 \ar@{-}[rr] \ar@{-}[dd] & \ar@{-}[d] &   IX^\dagger  \ar@{-}[dd]  \ar@{-}[rr] &  \ar@{-}[u]  & IX^\dagger &        
 \\%
&  XI  \ar@{-}[ld] & \ar@{-}[l] &  \ar@{-}[u] \ar@{-}[l]  \ar@{-}[r]  & \ar@{-}[r] &  XI   \\%
\ar@{-}[rr] &  &  \ar@{-}[ru]  & &  \ar@{-}[ll]  \ar@{-}[uu]  \ar@{-}[ur]                    %
}%
\end{array}.
\label{eq:Z3FSL}
\end{align}
where now $X$ and $Z$ are $\mathbb{Z}_3$ clock operators satisfying $XZ = e^{\frac{2\pi i}{3}}ZX$.
We have found that this model can be obtained by coupling together fractal cluster state layers~\cite{fractalSSPT,devakul2018universal,Devakul2018} stacked along the $\hat{z}$ and $\hat{y}$ directions, respectively. 
The Hamiltonians for the fractal cluster state layers are given by translates of the following local generators 
\begin{align}
\begin{array}{c}
\xymatrix@!0{%
  IZ \ar@{-}[dd]  \ar@{-}[rr]  & & \ar@{-}[rr] & & IZ
\\
 & X & & & 
\\
  IZ & &  \ar@{-}[ll]  \ar@{-}[uu]  & &  \ar@{-}[ll]  \ar@{-}[uu] 
}
\end{array},
\begin{array}{c}
\xymatrix@!0{%
   \ar@{.}[dd]  \ar@{.}[rr]  & & \ar@{.}[rr] & & Z
\\
 & & & IX & 
\\
  Z & &  \ar@{.}[ll]  \ar@{.}[uu]  & &  \ar@{.}[ll]  \ar@{.}[uu] Z
}
\end{array},
\nonumber
\end{align}
\begin{align}
\begin{array}{c}
\xymatrix@!0{%
&& ZI \ar@{-}[ddll] \ar@{-}[rr] &  & 
\\
& & X &
\\
 ZI &  & ZI  \ar@{-}[ll] \ar@{-}[uurr]
}
\end{array}
,
\begin{array}{c}
\xymatrix@!0{%
&& Z \ar@{.}[ddll] \ar@{.}[rr] &  & Z
\\
& & XI &
\\
  &  & Z  \ar@{.}[ll] \ar@{.}[uurr]
}
\end{array}.
\nonumber
\end{align}
where each vertex hosts two qubits and each $xy$ or $xz$ plaquette hosts a single qubit.
Here dotted lines indicate the dual \textit{square} lattices in $xy$ and $xz$ planes, respectively, rather than the dual of the cubic lattice. 
Each pair of terms on a line in the equation above generates a fractal cluster state on the sites of the square lattice and dual lattice in an $xy$ or $xz$ plane, respectively. 
We group pairs of spins on $xy$ and $xz$ plaquettes onto the adjacent corner with minimal $(x,y,z)$ coordinate, 
then in the limit of a strong $-(XX+ZZ^\dagger)$ field on these plaquette spins we find the FSL Hamiltonian in Eq.~\eqref{eq:Z3FSL} at leading order. 

We have found that the above coupled cluster state layer construction generalizes to all FSL models, see the appendix for further details. The general situation is depicted in Fig.~\ref{fig:FSL}, where the $xy$ fractal cluster layers in Fig.~\ref{fig:xzLayers} and $xz$ fractal cluster layers in Fig.~\ref{fig:xyLayers} are strongly coupled to produce the FSL model in Fig.~\ref{fig:xyzLayers}. 
In many cases (including the example above) the cluster layers can be further decomposed into coupled cluster chains.

\section{Gauging and cluster states} 
\label{sec:gauging}
\begin{figure}[t!]
    \centering
    \includegraphics[width=0.9\columnwidth]{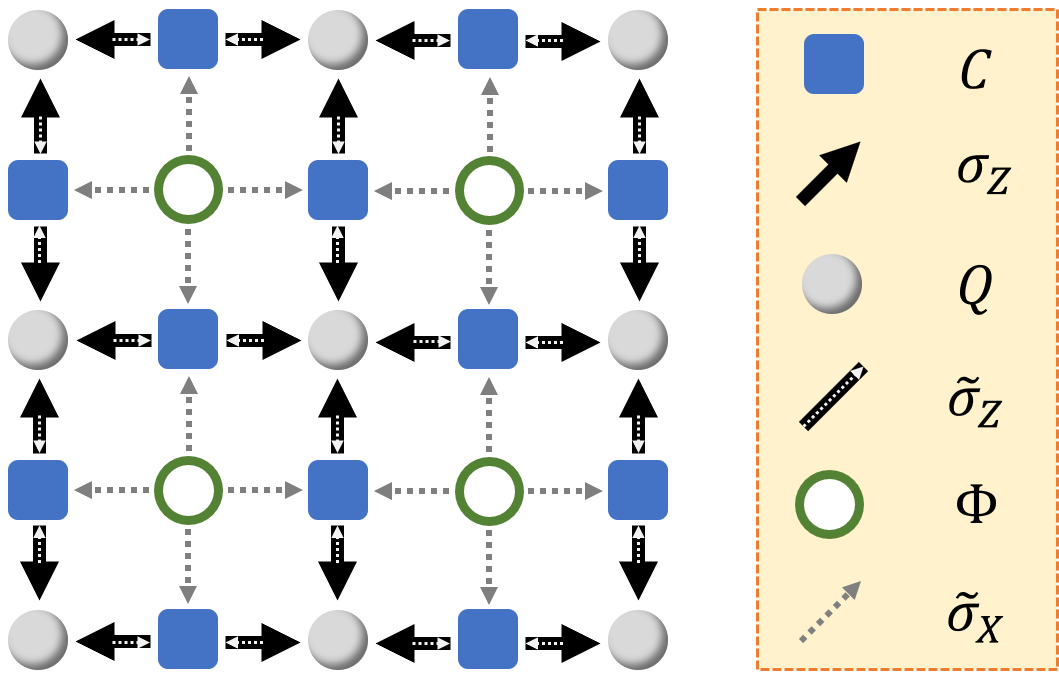}
    \caption{ 
    The figure on the left depicts gauging the 2D Ising model to obtain the 2D toric code. 
    The key on the right shows the constraints $C$, spins $Q$, local relations $\Phi$, and stabilizer map $\sigma_Z$ of the 2D Ising model. After gauging (and taking the $\alpha\rightarrow \infty$ limit described in the main text) $C$ labels qubits, $Q$ labels $Z$-type stabilizer generators, $\Phi$ labels $X$-type stabilizer generators, with $\widetilde{\sigma}_Z,\widetilde{\sigma}_X,$ the respective gauged stabilizer maps for the 2D toric code. 
}
\label{fig:2DToricCode}
\end{figure}

In this section we describe how certain cluster state models can be driven into topological phases by large on-site fields. In particular, we describe how to construct such a cluster state for any topological order that is obtained by gauging a subsystem symmetry following a generalized gauging procedure~\cite{vijay2016fracton,PhysRevB,shirley2018FoliatedFracton,kubica2018ungauging}.
Fig~\ref{fig:2DToricCode} shows this process applied to the 2D Ising model, resulting in the 2D toric code.

We define a \textit{locally specified} symmetry to be a group whose elements are products of Pauli $X$ matrices that commute with a set of local Pauli $Z$ constraints and whose composition rule is given by matrix multiplication. 
We utilize the stabilizer formalism~\cite{Gottesman1997} to describe the group of Pauli matrices under multiplication via linear algebra over $\mathbb{Z}_2$ ($\mathbb{Z}_d$ for qudits). 
In particular, a pair of basis vectors $\hat{s}_X,\hat{s}_Z$ are assigned to every spin $s\in Q$ to keep track of the exponents of $X_s$ and $Z_s$. 
An arbitrary Pauli operator (up to a $\pm$ sign) is then mapped to
\begin{align}
\prod_{s\in Q} X_s^{x(s)} Z_s^{z(s)}  \mapsto  \sum_{s \in Q} x(s) \hat{s}_X + z(s) \hat{s}_Z \, ,
\end{align}
where $x(s),z(s),$ are binary functions that store the power of Pauli matrices $X_s,Z_s,$ acting on spin $s$, respectively. 
The vector space describing all Pauli matrices $P$ breaks up into a direct sum of subspaces describing products of $X$ or $Z$ matrices, $P=P_X \oplus P_Z$, respectively. We point out that $P_X\cong P_Z \cong \mathbb{Z}_2[Q]$. 

We describe a set of Pauli $Z$ constraint terms, $\Gamma_c$, indexed by $c\in C$ via a linear map ${\sigma_Z: \mathbb{Z}_2[C] \rightarrow P_Z}$ defined such that 
\begin{align}
    \Gamma_c=\sigma_Z \hat{c} \, .
\end{align}
In the above equation we have promoted the constraint labels $c\in C$ to basis vectors $\hat{c}$ of $\mathbb{Z}_2[C]$. 
We remark that $\sigma_Z$ does not denote a single Pauli-$Z$ matrix, rather it is a linear map that sends basis vectors labelled by constraints $\hat{c}$ to $\mathbb{Z}_2$-linear combinations in $P_Z$ that represent products of Pauli-$Z$ matrices within the stabilizer formalism. 
Here we focus on cases where the $\Gamma_c$ are all local in two or three dimensional space.
In the example, Fig.~\ref{fig:2DToricCode}, $\sigma_Z$ maps a bond $c$ to the Ising $ZZ$ interaction term across that bond.
The violation, or excitation, of local constraints by Pauli $X$ matrices is described by the map ${\epsilon_Z: P_X \rightarrow \mathbb{Z}_2[C]}$, where 
\begin{align}
    \epsilon_Z =  \sigma_Z^\dagger \Lambda \, ,
\end{align} 
with symplectic bilinear form ${\Lambda: \hat{s}_X \leftrightarrow \hat{s}_Z}$.
This maps a Pauli $X$ operator to the set of bond constraint terms it anticommutes with.

The constraints $\sigma_Z$ determine a group of locally specified symmetries given by $\ker \epsilon_Z$. 
The locally specified symmetries can be further broken up into \textit{local} and \textit{global} symmetries by resolving as large a subspace of $\ker \epsilon_Z$ as possible via the image of a map ${\sigma_X: \mathbb{Z}_2[G_{\text{loc}}] \rightarrow P_X}$ that defines some local Pauli-$X$ terms,
\begin{align}
    \Sigma_g=\sigma_X \hat{g} \, ,
\end{align} 
indexed by a set of generators $g\in G_{\text{loc}}$. 
The local symmetries are then given by $\im \sigma_X \subseteq \ker \epsilon_Z$ and the global symmetries, modulo local symmetries, are given by $\ker \epsilon_Z / \im \sigma_X$.

Our starting point is a symmetric Hamiltonian in the paramagnetic phase 
\begin{align}
H = - \alpha \sum_{s} X_s - \sum_c \Gamma_c \, ,
\end{align}
in the limit $\alpha \gg 1$. 
This model can be gauged following the general procedure introduced in Ref.~\onlinecite{PhysRevB} for gauging locally specified symmetries\footnote{Here we use a slightly different convention to Ref.~\onlinecite{PhysRevB}, related by applying Hadamard gates to all gauge spins.
}. 
\begin{itemize}
    \item 
     The first step is to introduce a $\mathbb{Z}_2$ \textit{gauge spin} $c$ for each local constraint $\Gamma_c$. This extends the space of Pauli matrices to $P \oplus \widetilde{P}$ where $\widetilde{P}=\widetilde P_X\oplus\widetilde P_Z$ is the space of Pauli matrices on the gauge spins. We point out that $\widetilde P_X\cong \widetilde P_Z \cong \mathbb{Z}_2[C]$. 
    \item
     Next, we define gauge transformation operators $X_s A_s$, where $A_s = \widetilde\sigma_Z \hat{s}$ for $\widetilde\sigma_Z := \epsilon_Z$ considered as a map $\widetilde\sigma_Z :\mathbb{Z}_2[Q]\rightarrow \widetilde{P}_Z$ via the isomorphisms ${\mathbb{Z}_2[Q] \cong P_X ,}\, {\widetilde{P}_Z \cong \mathbb{Z}_2[C]}$, see Fig.~\ref{fig:2DToricCode}. 
    The original global symmetries can be recovered from products of the gauge transformation operators that act as identity on the gauge spins. 
    \item
     Gauge invariant states, i.e. those with eigenvalue $+1$ under all gauge transformations, satisfy a generalized Gauss's law. A projection onto this gauge invariant subspace is then introduced. Here we only enforce such a projection energetically by adding terms to the Hamiltonian that introduce an energy penalty $\geq \Delta$ for violation of Gauss's law. In the limit of $\Delta \rightarrow \infty$ the projection becomes strict. 
    \item
     Finally, the other Hamiltonian terms are modified to include a minimal coupling to the gauge spins by projecting them onto the gauge invariant subspace. For our purposes this simply amounts to the change ${\Gamma_c \mapsto X_c \Gamma_c}$. 
\end{itemize}

The gauged paramagnetic Hamiltonian is 
\begin{align}
\label{eq:ClusterHam}
H = - \alpha \sum_{s\in Q} X_s - \sum_{c \in C} X_c \Gamma_c - \Delta \sum_{s \in Q}   X_s A_s \, . 
\end{align}
This is a {cluster Hamiltonian} with a large $X$ field applied to the $s$ sites. 
See the section below for a further discussion of {cluster states}. 

We consider the limit $\alpha \rightarrow \infty$ with $\Delta$ held constant. 
This effectively fixes out all the original $Q$ spins in the $\ket{+}$ state. 
On the remaining $C$ spins the Gauss's law terms become $X_s A_s \mapsto A_s$ and only products of minimally coupled constraint fields $X_c \Gamma_c$ that act trivially on the $Q$ spins enter the effective Hamiltonian. 
These terms are all contained within $\ker \sigma_Z$, where we have made use of the isomorphism $\widetilde{P}_X\cong \mathbb{Z}_2[C]$ to write $\sigma_Z:\widetilde{P}_X\rightarrow P_Z$. 
In fact, all local terms in $\ker \sigma_Z$ appear at finite order in perturbation theory. 
Products of these terms are described by the image of a map $\widetilde{\sigma}_X: \mathbb{Z}_2[\Phi] \rightarrow \widetilde{P}_X$, satisfying $\im \widetilde{\sigma}_X \subseteq \ker \sigma_Z$, that resolves the maximal subspace of local terms in $\ker \sigma_Z$. 
In particular, only a set of local generators given by a choice of basis elements 
\begin{align}
    B_\phi = \widetilde{\sigma}_X \hat \phi\, ,
\end{align} 
for $\phi \in \Phi$, need to be included in the Hamiltonian. 
Other terms in $\im \widetilde{\sigma}_X$ are given by products of the generators $B_\phi$, and their inclusion only results in a shifting of the energy levels without changing the quantum phase of matter. 
We remark that the $B_\phi$ terms were introduced by hand in the equivalent gauging construction of Ref.~\onlinecite{PhysRevB}. 

In the example in Fig.~\ref{fig:2DToricCode}, gauge spins $c$ live on the edges of the square lattice.  
The operators in $\im \widetilde{\sigma}_X$ are closed contractible loops of $X$ operators, and $B_\phi$ is chosen to be the product of four gauge spins $X_c$ around the plaquette $\phi$. 
The effective Hamiltonian in the large $\alpha$ limit is simply the 2D toric code. 
Combining this observation with the cluster state splitting procedure, introduced in the following section, leads to the coupled cluster chain construction of the 2D toric code that was presented below Eq.~\eqref{eq:2DTC}.

More generally, the effective Hamiltonian in the ${\alpha\rightarrow \infty}$ limit is given by
\begin{align}
H =- \Delta \sum_s A_s - \sum_\phi B_\phi \, ,
\label{eq:GaugedHam}
\end{align}
which corresponds to the CSS~\cite{PhysRevA.54.1098,Steane2551} stabilizer map ${\widetilde{\sigma}={\widetilde{\sigma}_X\oplus \widetilde{\sigma}_Z} : \mathbb{Z}_2[\Phi] \oplus \mathbb{Z}_2[Q] \rightarrow \widetilde{P}}$. 
The Hamiltonian in Eq.~\eqref{eq:GaugedHam} is the same as the gauged and ``disentangled" Hamiltonian from Ref.~\onlinecite{PhysRevB}. 
Hence we have demonstrated that applying a large $X$ field to the $Q$ spins of the cluster Hamiltonian in Eq.~\eqref{eq:ClusterHam} results in the model obtained by gauging the paramagnetic phase under the locally specified symmetry $\ker \sigma_Z$. 

As an alternative to applying a large $X$ field to the $Q$ spins of the cluster Hamiltonian, we could instead measure them in the $X$ basis. This procedure prepares the gauged model in some excited eigenstate. A correction operator can then be found to restore the model to its groundstate~\cite{bravyi2013quantum,Brown2019}. See the section~\ref{sec:modular} for further discussion of this option.

\section{Coupling cluster state layers \& chains} 
\label{sec:CoupledLayers} 
In this section we describe a construction of certain cluster states by coupling together a number of different, lower dimensional, cluster states. When combined with the results from the previous section this leads to a coupled cluster state layer construction for certain stabilizer topological orders. 

A \textit{cluster Hamiltonian} on an arbitrary graph $\Theta$, with no self-loops or duplicate edges, is given by a sum of commuting Pauli terms 
\begin{align}
H_\Theta = - \sum_{v} X_v \prod_{v' \in N_v} Z_{v'} \, ,
\end{align}
where $v$ are vertices of the graph and $N_v$ is the set of nearest-neighbor vertices connected by an edge to $v$. 
The cluster Hamiltonian is in the trivial phase if no symmetries are enforced. 
This follows from the local unitary equivalence 
\begin{align}
U H_\Theta U^\dagger = - \sum_v X_v \, , && \text{where} && U= \prod_{\langle v v' \rangle} CZ_{v v'} \, ,
\label{eq:CZ}
\end{align}
where $\langle v v' \rangle$ denotes an edge in $\Theta$ between vertices $v,v',$ and $CZ$ is a controlled-$Z$ gate. 
This also reveals a simple construction of $H_{\Theta}$'s ground state $\ket{\psi_0} = U \ket{+}^{\otimes V}$
which is known as a \textit{cluster state}. 
On the other hand, each cluster state respects a (potentially nontrivial) symmetry group determined by the graph $\Theta$, which is given by Pauli-$X$ operators that are products of the local terms in the Hamiltonian. 
The cluster state may lie in a nontrivial (subsystem) symmetry-protected topological phase~\cite{chen2012symmetry,you2018subsystem,fractalSSPT} with respect to this symmetry. 

The Hamiltonian in Eq.~\eqref{eq:ClusterHam} above is a cluster Hamiltonian on the \textit{interaction graph} of the constraints $\sigma_Z$, under a large $X$ field on $s$ sites. 
The interaction graph of a set of constraints is a bipartite graph with a vertex for each spin and each constraint, with an edge connecting each constraint to the spins it acts nontrivially on.

\begin{figure}[t!]
    \centering
    \subfloat[\label{fig:graph1}]{\includegraphics[height=0.25\columnwidth]{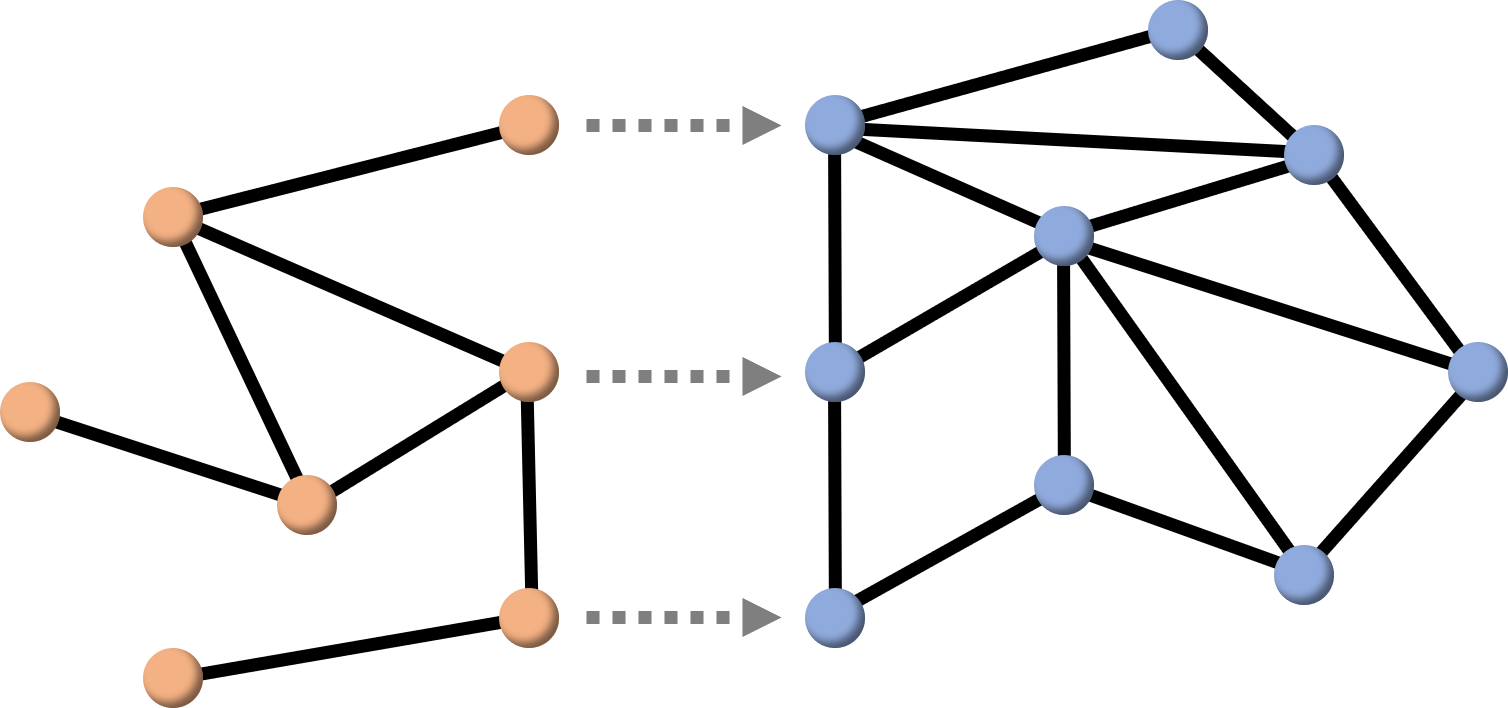}}
    \hspace{.2cm}
    \subfloat[\label{fig:graph2}]{\includegraphics[height=0.25\columnwidth]{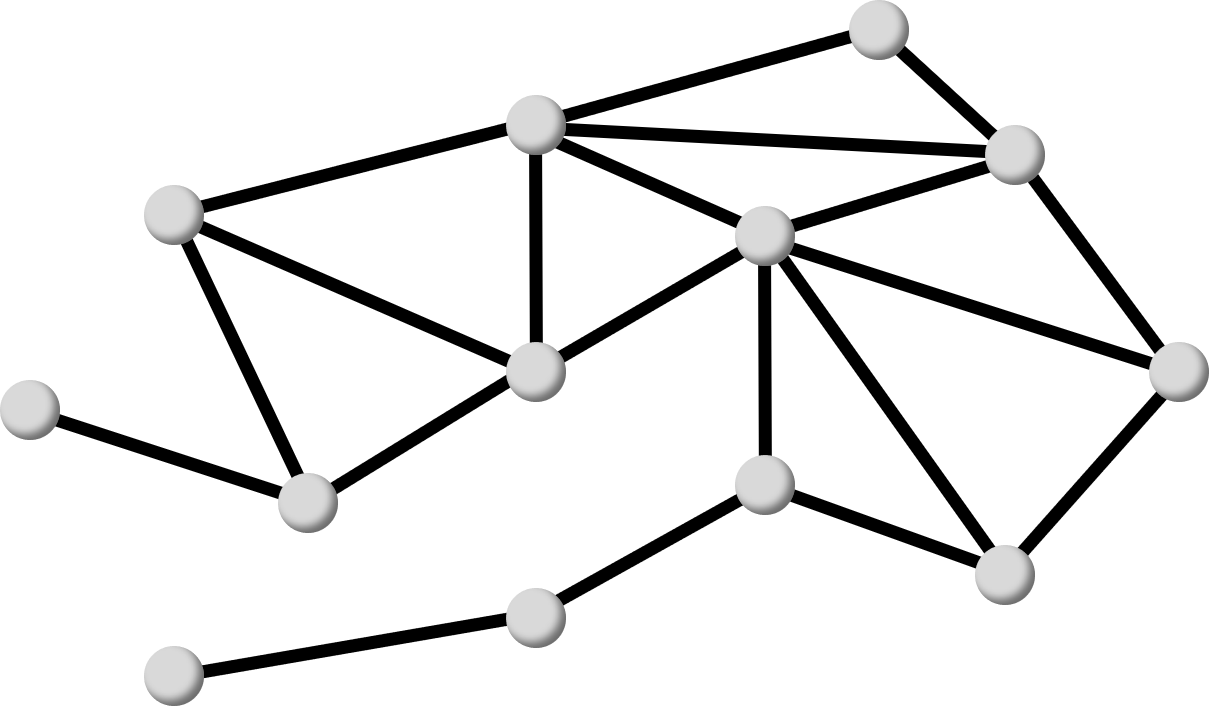}}
    \caption{
    (a) An example of graphs $\Theta$ (orange) and $\Theta'$ (blue) and a gluing map $k$ (arrows) involving a subset of vertices. 
    (b) The graph $\Theta \cup_k \Theta'$ obtained by gluing along $k$. 
}
\end{figure}

To obtain a coupled layer construction for a gauged Hamiltonian, as in Eq.~\eqref{eq:ClusterHam}, we make use of an elementary \textit{merging} and \textit{splitting} property of cluster Hamiltonians. 
Two separate cluster Hamiltonians, one containing a vertex $v$ and the other a $v'$, can be strongly coupled via a $Z_v Z_{v'}$ field to produce a single cluster Hamiltonian on a new graph where $v$ and $v'$ have been identified. 
More precisely, cluster Hamiltonians on the graphs $\Theta$ and $\Theta'$ can be coupled together to produce a model equivalent to a cluster Hamiltonian on a new graph $\Theta \cup_{k} \Theta'$ in the strong coupling limit. 
Here $k$ denotes an injective map from a subset $\lambda$ of the vertices of $\Theta$ to the vertices of $\Theta'$, see Fig.~\ref{fig:graph1}. 
The graph $\Theta \cup_{k} \Theta'$ is obtained by \textit{gluing} $\Theta$ to $\Theta'$, i.e. identifying $v$ with $k(v)$ for all $v\in\lambda$, see Fig.~\ref{fig:graph2}. 
In the gluing process edges are added modulo 2, meaning that any doubled edges are deleted, i.e. if $\langle u,v\rangle\in \Theta$ and $\langle k(u),k(v) \rangle \in \Theta'$ then there is no $\langle u,v \rangle$ edge in $\Theta \cup_{k} \Theta'$, for any $u,v\in\lambda$.

The coupled Hamiltonian is given by 
\begin{align}
H(\alpha)=H_\Theta + H_{\Theta'} - \alpha \sum_{v \in \lambda} Z_v Z_{k(v)}  \, .
\end{align}
In the $\alpha \rightarrow \infty$ limit the two qubit Hilbert space on vertices $v$ and $k(v)$ is projected into the single qubit subspace with effective Pauli operators $X_v X_{k(v)} \mapsto \overline{X}_v$ and $Z_v \sim Z_{k(v)} \mapsto \overline{Z}_v$, where $\sim$ indicates that these operators become identified within the subspace. 
In this limit the Hamiltonian becomes
\begin{align}
H(\alpha) & =  - \sum_{v \in \Theta \backslash \lambda} X_v \prod_{u \in N_v} Z_{u} 
- \sum_{v' \in \Theta' \backslash k(\lambda)} X_{v'} \prod_{u' \in N_{v'}} Z_{u'}
\nonumber \\
&-  \sum_{v \in \lambda} X_v X_{k(v)} \prod_{u \in N_v} Z_{u} \prod_{u' \in N_{k(v)}} Z_{u'}  - \alpha \sum_{v \in \lambda} Z_v Z_{k(v)}
\, ,
\nonumber
\end{align}
at leading order in perturbation theory as $\alpha \rightarrow \infty$ and hence $H(\infty) \mapsto  \ H_{\Theta \cup_k \Theta'} $. 

\begin{figure}[t!]
    \centering
    \subfloat[\label{fig:cluster2D}]{\includegraphics[width=0.43\columnwidth]{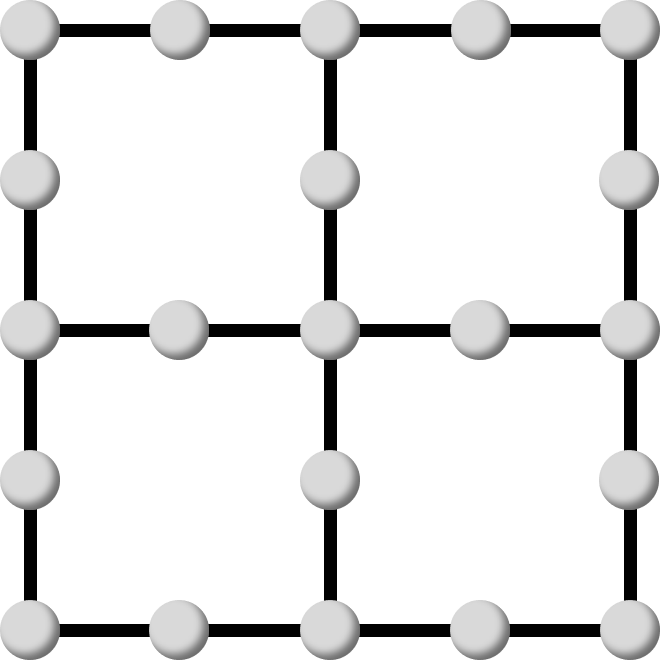}}
    \hspace{.5cm}
    \subfloat[\label{fig:clusterchains}]{\includegraphics[width=0.49\columnwidth]{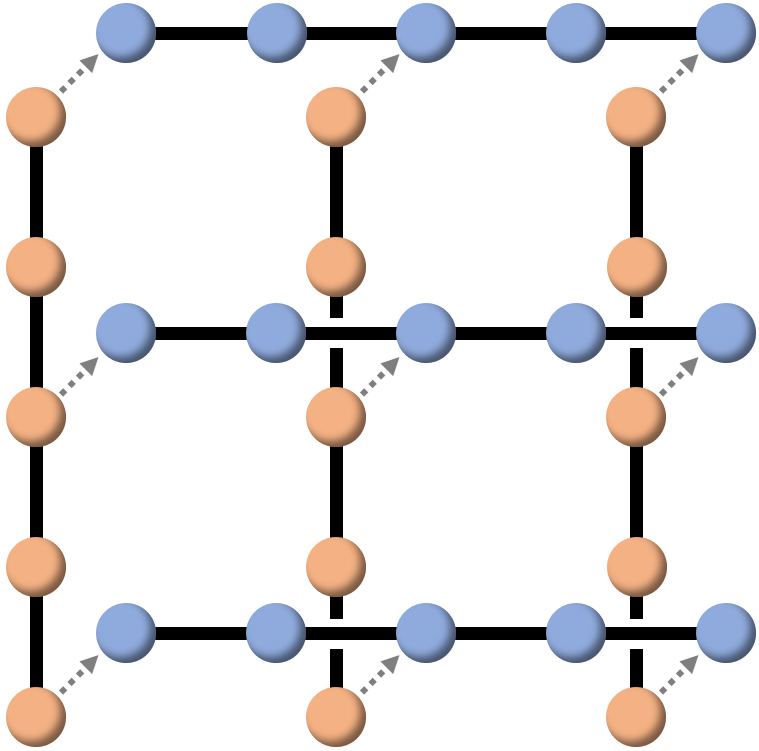}}
    \caption{
    (a) The cluster state corresponding to the gauged Ising model in Fig.~\ref{fig:2DToricCode}.
    (b) A construction of (a) from coupled 1D cluster chains. 
}
\end{figure}

This implies that any cluster Hamiltonian $H_\Theta$ can be recovered by coupling together cluster Hamiltonians $H_{\theta_i}$ on a set of $N$ subgraphs $\theta_i$ that cover $\Theta$, with $\mathbb{Z}_2$ edge addition, i.e. $\Theta =  { \theta_1 \cup \theta_2 \cup \dots \cup \theta_N} $. 
For appropriate graphs $\Theta$ this can result in a coupled layer or chain construction. 
For example, the cluster state on the vertices and edges of a grid, Fig.~\ref{fig:cluster2D}, is given by coupled cluster chains, Fig.~\ref{fig:clusterchains}. 
As discussed in the section above, applying a large $X$ field to all vertices of the grid in Fig.~\ref{fig:cluster2D} yields the coupled cluster chain construction of the toric code introduced below Eq.~\eqref{eq:2DTC}. 

More generally, combining the coupled cluster state construction from this section with the gauged paramagnet limit of a cluster Hamiltonian under a large $X$ field in Eq.~\eqref{eq:ClusterHam} yields a coupled cluster state construction of any gauged paramagnet Hamiltonian. 
The gauged Hamiltonian, Eq.~\eqref{eq:GaugedHam}, with interaction graph $\Theta=\cup_i \theta_i$ is obtained in the $\alpha \rightarrow \infty$ limit from
\begin{align}
\sum_{i \text{-layer}} H_{\theta_i}
-\alpha \sum_{v \in \Theta} \sum_{u\neq u' \in \Omega_v} Z_{u}Z_{u'} - \alpha \sum_{s\in Q} \prod_{u\in \Omega_s} X_u
\, .
\label{eq:generalcoupledcluster}
\end{align}
Where $s$ runs over the $Q$ vertices of $\Theta$ that are being projected out in Eq.~\eqref{eq:GaugedHam} and $\Omega_v$ denotes the set of vertices in different subgraphs $\theta_i$ that become identified with $v$ in $\Theta$.

The largest degree interaction in the coupled cluster state Hamiltonian above is determined by the maximum degree of the cluster terms, which is one greater than the maximum degree of all the subgraphs $\theta_i$, and the maximum degree of the $\prod X$ fields, which is the largest number of different subgraphs $\theta_i$ sharing a vertex that becomes identified with a $Q$ vertex in $\Theta$ which gets projected out. 
In particular, this leads to many coupled chain and coupled layer constructions of 2D and 3D topological stabilizer Hamiltonians with low interaction weight. 
This construction applies equally well to Hamiltonians with conventional topological order, type-I, and type-II fracton topological order, as demonstrated by the examples above and in the appendix. 

\section{Modular architecture implementation} 
\label{sec:modular}

In this section we describe an implementation of our coupled spin chain, and layer, constructions in modular architectures~\cite{Thaker2006,Jiang2007,Moehring2007,Helmer2009,Yao2010,Fujii2012,Monroe2012} that combine static qubits arranged into 1D or 2D subsystems,  such as ion traps or superconducting qubits, with auxiliary degrees of freedom that are not strictly constrained by locality, for example photonic qubits.  

In the modular architecture the auxiliary qubits are those in $\{ \Omega_s \}_{s\in Q}$  from Eq.~\eqref{eq:generalcoupledcluster}, while the remaining qubits in each graph $\theta_i$ are stored in the static isolated subsystems with one or two dimensional locality. 
An additional auxiliary qubit, labelled by $\langle u,u'\rangle$, is introduced for each $Z_uZ_{u'}$ operator in Eq.~\eqref{eq:generalcoupledcluster} that is not contained within $\{ \Omega_s \}_{s\in Q}$, to facilitate the measurement of this operator. The qubits in the static subsystems persist and hold the state of the desired stabilizer model, while the auxiliary qubits are entangled with the static subsystem qubits and then destructively measured to implement the desired stabilizer measurements on the static qubits.

The ground state of the gauged Hamiltonian in Eq.~\eqref{eq:GaugedHam} is prepared by initializing all qubits in the $\ket{+}$ state and then entangling them via CZ gates to create cluster states on the graphs $\theta_i$. 
Next, a pair of CZ gates is applied to entangle each additional auxiliary qubit $\langle u,u'\rangle$, introduced above, with the qubits $u$ and $u'$. 
The $Z_uZ_{u'}$ and $\prod X_u$ operators from Eq.~\eqref{eq:generalcoupledcluster} are then measured on the auxiliary qubits, projecting them into a unique state, decoupled from the static qubits. 
For the additional auxiliary qubits $\langle u,u'\rangle$, this will involve a measurement in the $X$ basis to implement the desired $Z_uZ_{u'}$ measurements on static qubits $u$ and $u'$. 
If the stabilizer measurements all return the $+1$ eigenvalue, the desired ground state has been prepared. 
If any of the $Z_uZ_{u'}$ stabilizer measurements return $-1$ eigenvalues we can find a Pauli-$X$ correction operator supported on the static qubits to flip these eigenvalues back to $+1$. 
If any of the $\prod_{u\in\Omega_s} X_u$ stabilizer measurements returns a $-1$ eigenvalue, we have effectively projected the static qubits into the $-1$ eigenvalue of the associated $A_s$ operator. 
For topological codes~\cite{Dennis2001}, including fracton codes~\cite{bravyi2013quantum,Brown2019}, we can decode these errors to find a Pauli-$X$ correction operator. 
After applying the product of the correction operators, we have prepared a state within the desired ground space.

We remark that the stabilizer generators for the decoupled cluster states together with the $Z_uZ_{u'}$ and $\prod X_u$ operators generate a subsystem stabilizer code~\cite{bacon,poulin} (which does not encode a logical qubit). In this context, measurement of the $Z_uZ_{u'}$ and $\prod X_u$ operators, followed error-correction, corresponds to gauge fixing~\cite{Paetznick2013} onto the desired topological code.

More generally, measurement of the $A_s$ stabilizers on the static qubits can be performed following the above method. 
Measurement of the $B_{\phi}$ stabilizers can be accomplished similarly, by introducing a dual set of auxiliary qubits. 
Provided the rate of errors between measurement and correction steps is below the threshold for the code~\cite{Dennis2001,bravyi2013quantum,Brown2019} the measurement and correction process will successfully preserve logical information encoded into the ground space of the (fracton) topological code on the static qubits with high probability for sufficiently large system sizes. 
Hence for topological codes that admit a coupled layer construction, the procedure outlined in this section provides an implementation of topological error correction in a modular architecture, featuring low dimensional static subsystems coupled to long range auxiliary systems. 


\section{Discussion \& conclusions.} 
\label{sec:conclusion}
In this paper we have introduced coupled spin chain, and layer, constructions for topological stabilizer models, including conventional, type-I fracton, and type-II fracton topological orders. 
We based our construction on two key properties of cluster states. 
First, projecting out an extensive number of spins with a large $X$ field produces a gauged model on the remaining spins. 
Second, many cluster states naturally split into coupled lower dimensional cluster states. 
In particular, this provides a coupled cluster chain construction of Haah's cubic code and a coupled cluster layer construction of Yoshida's FSL models. 

Our coupled layer constructions involve relatively low degree Hamiltonian interactions, thus paving the way towards experimental realization of exotic fracton topological orders. 
They are particularly well suited to modular architectures where qubits are organized into subsystems with inherent one or two dimensional local interactions. 
These subsystems need only interact with nonlocal auxiliary degrees of freedom that can be jointly measured for our scheme to be realizable, as explained above. 
Similar schemes have recently appeared in the context of photonic topological quantum computing~\cite{Bartolucci2021,Bombin2021}, it would be interesting to recast our constructions in this setting.

Looking forward, it would be interesting to find a coupled layer construction of more general, possibly nonabelian, fracton models. 
We remark that our construction provides a hint in this direction, as the merging and splitting of cluster states also applies to generalized cluster states where the $Z$ operators are replaced by other diagonal operators, such as $\sqrt{Z}$, $CZ$ or multiply controlled-$Z$~\cite{levin2012braiding,Ni2014,Fidkowski2019}. 
It would also be very interesting to derive topological defect networks from our coupled layer constructions.

\acknowledgements
We acknowledge useful discussions with A.~T. Schmitz. 
DW thanks Arpit Dua for a useful conversation about gauge fixing. 
DW acknowledges support from the Simons foundation. 
TD acknowledges support from the Charlotte Elizabeth Procter Fellowship at Princeton University.


\bibliographystyle{apsrev4-1}
\bibliography{refs}

\clearpage 

\appendix

\onecolumngrid

\section{Further coupled layer examples}
\label{app:examples}

In this section we apply our coupled layer construction to a number of further examples including the 3D toric code~\cite{Dennis2001}, the X-cube model~\cite{vijay2016fracton}, FSL models~\cite{yoshida2013exotic}, and cubic code B~\cite{haah2014bifurcation}. 
These examples are presented in terms of local stabilizer generators $\{h_i\}_i$ that appear in the commuting projector Hamiltonian $H = - \sum_i h_i$ whose ground space is the stabilizer codespace. 
Specifically, we first present the stabilizer generators of the cluster chains or layers to be coupled. We then describe the on-site couplings between the layers. Finally the stabilizer generators of the coupled Hamiltonian, in the limit of infinite coupling field strength, are given.  

%

\subsection{3D toric code}

In this subsection we describe constructions of 3D toric code from 1D or 2D cluster states. 

First we describe a construction of the 3D toric code from coupled 1D cluster states. The cluster chains run along the $\hat{x},\hat{y},\hat{z}$ axes of a cubice lattice, with one qubit per edge and three qubits per vertex. The stabilizer generators are given by 
\begin{align}
\begin{array}{c}
\xymatrix@!0{%
& &
\\
Z   & XII  \ar@{-}[u]  \ar@{-}[d]  \ar@{-}[l]  \ar@{-}[r] \ar@{-}[ur] \ar@{-}[dl]     & Z
\\
&
}
\end{array},
&&
\begin{array}{c}
\xymatrix@!0{%
ZII   & X  \ar@{-}[l] \ar@{-}[r]  & ZII
}
\end{array},
&&
\begin{array}{c}
\xymatrix@!0{%
& Z &
\\
& IXI \ar@{-}[u]  \ar@{-}[d]  \ar@{-}[l]  \ar@{-}[r] \ar@{-}[ur] \ar@{-}[dl]  & 
\\
& Z
}
\end{array},
&&
\begin{array}{c}
\xymatrix@!0{%
 IZI 
\\
 X  \ar@{-}[u]  \ar@{-}[d]   &
\\
 IZI
}
\end{array},
&&
\begin{array}{c}
\xymatrix@!0{%
& & Z 
\\
& IIX \ar@{-}[u]  \ar@{-}[d]  \ar@{-}[l]  \ar@{-}[r] \ar@{-}[ur] \ar@{-}[dl]   & 
\\
Z &&
}
\end{array},
&&
\begin{array}{c}
\xymatrix@!0{%
& &  II Z
\\
&  X  \ar@{-}[ur]  \ar@{-}[dl]   &
\\
 IIZ &
}
\end{array}.
\label{eq:3D1DCS}
\end{align}
A strong $-(XXX+ZZI+ZIZ+IZZ)$ coupling on all vertices leads to the local stabilizer generators of the 3D toric code on the cubic lattice with one qubit per edge 
\begin{align}
\begin{array}{c}
\xymatrix@!0{%
& Z & Z
\\
Z &  \ar@{-}[u]  \ar@{-}[d]  \ar@{-}[l]  \ar@{-}[r] \ar@{-}[ur] \ar@{-}[dl]  & Z
\\
Z & Z
}
\end{array},
&&
\begin{array}{c}
\xymatrix@!0{%
  & X \ar@{-}[l]  \ar@{-}[r]  & 
\\
 X  \ar@{-}[u]  \ar@{-}[d]   & & X \ar@{-}[u]  \ar@{-}[d]
\\
  & X \ar@{-}[l]  \ar@{-}[r] & 
}
\end{array},
&&
\begin{array}{c}
\xymatrix@!0{%
&& 
\\
& X \ar@{-}[dl] \ar@{-}[ur]  & X \ar@{-}[u] \ar@{-}[d]
\\
 &&
\\
X  \ar@{-}[u]\ar@{-}[d]  & X \ar@{-}[dl] \ar@{-}[ur]
\\
&&
}
\end{array},
&&
\begin{array}{c}
\xymatrix@!0{%
&& & X  \ar@{-}[l]\ar@{-}[r] & 
\\
& X \ar@{-}[dl] \ar@{-}[ur]  & & X \ar@{-}[ur] \ar@{-}[dl] 
\\
  & X  \ar@{-}[l]\ar@{-}[r] & 
}
\end{array}.
\end{align}
A further generalization to $d$-dimensional toric code from coupled 1D cluster states along $d$ different axial directions is straightforward. 

The above construction of the $d$-dimensional toric code with qubits on edges from coupled 1D cluster states can be extended to a construction of toric code with qubits on $k$-cells, for $1<k<d$, from coupled $k$-dimensional cluster states with qubits on $(k-1)$- and $k$-cells. 
For example, 3D toric code can be constructed from coupled 2D edge-face cluster states on the cubic lattice with 1 qubit per face and two qubits per edge. The stbailizer generators are 
\begin{align}
\begin{array}{c}
\xymatrix@!0{%
  & IZ \ar@{-}[l]  \ar@{-}[r]  & 
\\
 IZ  \ar@{-}[u]  \ar@{-}[d]   & X & IZ \ar@{-}[u]  \ar@{-}[d]
\\
  & IZ \ar@{-}[l]  \ar@{-}[r] & 
}
\end{array},
&&
\begin{array}{c}
\xymatrix@!0{%
  &  \ar@{-}[l]  \ar@{-}[r]  & & \ar@{-}[l]  \ar@{-}[r] & 
\\
   \ar@{-}[u]  \ar@{-}[d]   & Z & IX \ar@{-}[u]  \ar@{-}[d] & Z & \ar@{-}[u]  \ar@{-}[d]
\\
  &  \ar@{-}[l]  \ar@{-}[r] & & \ar@{-}[l]  \ar@{-}[r] & 
}
\end{array},
&&
\begin{array}{c}
\xymatrix@!0{%
  & \ar@{-}[l]  \ar@{-}[r]  & 
\\
  \ar@{-}[u]  \ar@{-}[d]   & Z &  \ar@{-}[u]  \ar@{-}[d]
\\
  & IX \ar@{-}[l]  \ar@{-}[r] & 
\\
  \ar@{-}[u]  \ar@{-}[d]   & Z &  \ar@{-}[u]  \ar@{-}[d]
\\
  &  \ar@{-}[l]  \ar@{-}[r] & 
}
\end{array},
\\
\begin{array}{c}
\xymatrix@!0{%
&& 
\\
& ZI \ar@{-}[dl] \ar@{-}[ur]  & ZI \ar@{-}[u] \ar@{-}[d]
\\
 & X &
\\
ZI  \ar@{-}[u]\ar@{-}[d]  & ZI \ar@{-}[dl] \ar@{-}[ur]
\\
&&
}
\end{array},
&& 
\begin{array}{c}
\xymatrix@!0{%
&& 
\\
&  \ar@{-}[dl] \ar@{-}[ur]  &  \ar@{-}[u] \ar@{-}[d]
\\
 & Z &
\\
\ar@{-}[u]\ar@{-}[d] & XI  \ar@{-}[dl] \ar@{-}[ur]  &  \ar@{-}[u] \ar@{-}[d]
\\
 & Z &
\\
  \ar@{-}[u]\ar@{-}[d]  &  \ar@{-}[dl] \ar@{-}[ur]
\\
&&
}
\end{array},
&&
\begin{array}{c}
\xymatrix@!0{%
& & & & 
\\
& & &  \ar@{-}[dl] \ar@{-}[ur]  &  \ar@{-}[u] \ar@{-}[d]
\\
 & & & Z & 
\\
&  \ar@{-}[dl] \ar@{-}[ur]  &  XI \ar@{-}[u] \ar@{-}[d] &  \ar@{-}[dl] \ar@{-}[ur] & 
\\
 & Z & & & 
\\
  \ar@{-}[u]\ar@{-}[d]  &  \ar@{-}[dl] \ar@{-}[ur] & & 
\\
& & & & 
}
\end{array},
\\
\begin{array}{c}
\xymatrix@!0{%
&& & ZI \ar@{-}[l]\ar@{-}[r] & 
\\
& IZ \ar@{-}[dl] \ar@{-}[ur]  & X & IZ \ar@{-}[ur] \ar@{-}[dl] 
\\
  & ZI  \ar@{-}[l]\ar@{-}[r] & 
}
\end{array},
&& 
\begin{array}{c}
\xymatrix@!0{%
&& &   \ar@{-}[l]\ar@{-}[r] & & \ar@{-}[l]\ar@{-}[r] & 
\\
&  \ar@{-}[dl] \ar@{-}[ur]  & Z & IX \ar@{-}[ur] \ar@{-}[dl] & Z & \ar@{-}[ur] \ar@{-}[dl]
\\
  &   \ar@{-}[l]\ar@{-}[r] & & \ar@{-}[l]\ar@{-}[r] & 
}
\end{array},
&&
\begin{array}{c}
\xymatrix@!0{%
& & & & &  \ar@{-}[l]\ar@{-}[r] & 
\\
& & &  \ar@{-}[dl] \ar@{-}[ur]  & Z &  \ar@{-}[ur] \ar@{-}[dl] 
\\
& &  & XI  \ar@{-}[l]\ar@{-}[r] & 
\\
&  \ar@{-}[dl] \ar@{-}[ur]  & Z &  \ar@{-}[ur] \ar@{-}[dl] 
\\
  &   \ar@{-}[l]\ar@{-}[r] & 
}
\end{array}. 
\end{align}
A strong $-(XX+ZZ)$ coupling on all edges leads to the local stabilizer generators of the 3D toric code on the dual cubic lattice, with one qubit per face 
\begin{align}
\begin{array}{c}
\xymatrix@!0{%
&& &  & \ar@{-}[lldd]\ar@{-}[dd]\ar@{-}[ll] 
\\
&   & X   &  &  
\\
 \ar@{-}[uurr]\ar@{-}[rr]\ar@{-}[dd]  & &   & X   & 
\\
  & X  & & & 
\\
 & & \ar@{-}[uurr]\ar@{-}[ll]\ar@{-}[uu]  & & 
}
\end{array},
\end{align}
\begin{align}
\begin{array}{c}
\xymatrix@!0{%
& & & & \ar@{-}[ll]\ar@{-}[rr]\ar@{-}[dd]  &  & 
\\
& & & Z \ar@{.}[ur] \ar@{.}[dl] &  \ar@{-}[u] &  
\\
& & \ar@{-}[ddll]\ar@{-}[rr]\ar@{-}[uu] \ar@{.}[d] &   &  \ar@{-}[dd] \ar@{-}[uurr] 
\\
& & Z \ar@{.}[d] &   &
\\
  &   & \ar@{-}[uurr]\ar@{-}[rr]\ar@{-}[ll]  &   & 
}
\end{array},
&&
\begin{array}{c}
\xymatrix@!0{%
& & & & 
\\
& & &   &  
\\
 & & \ar@{-}[uurr]\ar@{-}[dd]\ar@{-}[ddll]\ar@{-}[ll]\ar@{-}[rr]  & & \ar@{-}[uu]\ar@{-}[dd] 
\\
&  &  &  Z \ar@{.}[dl] \ar@{.}[ur] & 
\\
\ar@{-}[uu]\ar@{-}[dd]   & Z \ar@{.}[l] \ar@{.}[r] & \ar@{-}[rr]\ar@{-}[ddll] & & 
\\
   &   & & 
\\
& & & & 
}
\end{array},
&&
\begin{array}{c}
\xymatrix@!0{%
& & & & 
\\
& & &  &  
\\
& & \ar@{-}[uurr] \ar@{-}[ddll] & Z \ar@{.}[l] \ar@{.}[r] & \ar@{-}[uu] \ar@{-}[rr] & & 
\\
& & \ar@{-}[u]\ar@{-}[d] &  & Z \ar@{.}[u] \ar@{.}[d] & & 
\\
& &  \ar@{-}[dd]\ar@{-}[uu]\ar@{-}[uurr]\ar@{-}[ll]\ar@{-}[rr]  &  & \ar@{-}[ddll]\ar@{-}[uurr]
\\
& & &  
\\
& & & &
}
\end{array}. 
\end{align}

\subsection{X-cube model}

In this subsection we describe constructions of X-cube from coupled 2D or 1D cluster states. 

First we describe a construction from coupled 2D vertex-face cluster states on the cubic lattice with one qubit per face and three qubits per vertex. The local stabilizer generators of the cluster states are  
\begin{align}
\begin{array}{c}
\xymatrix@!0{%
  IIZ  \ar@{-}[dd]  \ar@{-}[rr]  & & IIZ
\\
 & X & 
\\
  IIZ & & IIZ \ar@{-}[ll]  \ar@{-}[uu] 
}
\end{array},
&&
\begin{array}{c}
\xymatrix@!0{%
&& ZII
\\
&   & 
\\
ZII \ar@{-}[dd]  \ar@{-}[uurr]  & X & ZII \ar@{-}[uu]  \ar@{-}[ddll] 
\\
 & 
\\
ZII &&
}
\end{array},
&& 
\begin{array}{c}
\xymatrix@!0{%
&& IZI \ar@{-}[ddll] \ar@{-}[rr] &  & IZI
\\
& & X &
\\
 IZI &  & IZI \ar@{-}[ll] \ar@{-}[uurr]
}
\end{array},
\end{align}
\begin{align}
\begin{array}{c}
\xymatrix@!0{%
  \ar@{-}[dd]  \ar@{-}[rr]   &   &  \ar@{-}[dd]  \ar@{-}[rr]   &&
\\
   & Z &  & Z &
\\
   \ar@{-}[dd]  \ar@{-}[rr]  & & IIX  \ar@{-}[dd]  \ar@{-}[rr]  & & \ar@{-}[dd]  \ar@{-}[uu]
\\
  & Z &  & Z &
\\
  &  &   \ar@{-}[ll]  \ar@{-}[rr]  & & 
}
\end{array},
&& 
\begin{array}{c}
\xymatrix@!0{%
& & & &  \ar@{-}[dd] \ar@{-}[ddll]
\\
& & & &
\\
 & & & Z & 
\\
&  &  & & 
\\
\ar@{-}[dd] \ar@{-}[uurr] & Z & XII  \ar@{-}[ddll] \ar@{-}[uurr] \ar@{-}[uu] \ar@{-}[dd] & Z & \ar@{-}[ddll] \ar@{-}[uu]
\\
& & & & 
\\
& Z & & & 
\\
& & & &
\\
 \ar@{-}[uu] \ar@{-}[uurr] & & & &
}
\end{array},
&&
\begin{array}{c}
\xymatrix@!0{%
& & && \ar@{-}[rr] \ar@{-}[ddll] &  & & & \ar@{-}[ll] \ar@{-}[ddll]
\\
& & & & Z & & Z & 
\\
&& & & IXI \ar@{-}[ll] \ar@{-}[rr] \ar@{-}[uurr] \ar@{-}[ddll] &  & 
\\
&   & Z & & Z & 
\\
\ar@{-}[uurr] \ar@{-}[rr]  & & & & \ar@{-}[ll] \ar@{-}[uurr]
}
\end{array}.
\end{align}
A strong $-(XX+ZZ)$ coupling on all edges leads to the local stabilizer generators of the X-cube model on the dual cubic lattice, with one qubit per face, in the limit of infinite field strength. The $Z$-type local stabilizer generator that appear at leading order are 
\begin{align}
\begin{array}{c}
\xymatrix@!0{%
& & & & & & & & 
\\
& & & & &   &  & & 
\\
& &  & & \ar@{-}[uurr]\ar@{-}[dd]\ar@{-}[ddll]\ar@{-}[ll]\ar@{-}[rr]  & & \ar@{-}[uu]\ar@{-}[dd] & & \ar@{-}[ll]\ar@{-}[ddll]
\\
& & &  &  &  Z \ar@{.}[dl] \ar@{.}[ur] & & & 
\\
& & \ar@{-}[uu]\ar@{-}[dd]   & Z \ar@{.}[l] \ar@{.}[r] & \ar@{-}[rr]\ar@{-}[ddll] & & \ar@{-}[dd]  & & 
\\
& &  & & Z \ar@{.}[u] \ar@{.}[d] & & & & 
\\
\ar@{-}[rr] \ar@{-}[uurr]  & & & &\ar@{-}[ddll]\ar@{-}[rr] \ar@{-}[ll] \ar@{-}[uurr]  & & 
\\
& &  &   & & & & & 
\\
& & \ar@{-}[uu] & & & & & & 
}
\end{array}.
\end{align}
The above term corresponds to the product $\prod_{p \ni v} Z_p$ of $Z_p$ on the twelve plaquettes that share a common vertex. Only three of the $Z_p$ terms have been depicted explicitly for clarity of the diagram. 

The $X$-type local stabilizer generators that appear at leading order in perturbation theory at infinite coupling field strength are
\begin{align}
\begin{array}{c}
\xymatrix@!0{%
&& &  & \ar@{-}[lldd]\ar@{-}[dd]\ar@{-}[ll] 
\\
&   & X   &  &  
\\
 \ar@{-}[uurr]\ar@{-}[rr]\ar@{-}[dd]  & &   &   & \ar@{-}[l]
\\
  & X  & & & 
\\
 & & \ar@{-}[uurr]\ar@{-}[ll]\ar@{-}[uu]  & & 
}
\end{array},
&&
\begin{array}{c}
\xymatrix@!0{%
&& \ar@{-}[d] &  & \ar@{-}[lldd]\ar@{-}[dd]\ar@{-}[ll] 
\\
&   &  &  &  
\\
 \ar@{-}[uurr]\ar@{-}[rr]\ar@{-}[dd]  & &   & X   & 
\\
  & X  & & & 
\\
 & & \ar@{-}[uurr]\ar@{-}[ll]\ar@{-}[uu]  & & 
}
\end{array},
&&
\begin{array}{c}
\xymatrix@!0{%
&& &  & \ar@{-}[lldd]\ar@{-}[dd]\ar@{-}[ll] 
\\
&   & X   &  &  
\\
 \ar@{-}[uurr]\ar@{-}[rr]\ar@{-}[dd]  & &   & X   & 
\\
  &  & & & 
\\
\ar@{-}[ur] & & \ar@{-}[uurr]\ar@{-}[ll]\ar@{-}[uu]  & & 
}
\end{array}.
\end{align}
Similar to the $Z$-type term above, these $X$-type terms depict $\prod_{p\in c,p\not \perp \hat{i}}X_p$ with $i=x,y,z,$ respectively. Only two of the $X_p$ terms are shown explicitly for clarity of the diagram. 

Alternatively we also find a construction of X-cube from coupled 1D cluster states along the axes of the cubic lattice, with one qubit per edge and three qubits per vertex. 
Starting from the stabilizer generators in Eq.~\eqref{eq:3D1DCS} and adding a strong $-(ZZZ+XXI+XIX+IXX)$ coupling to all vertices leads to  the stabilizer generators of the X-cube model on the cubic lattice with one qubit per edge in the infinite coupling strength limit
\begin{align}
\begin{array}{c}
\xymatrix@!0{%
& Z & 
\\
Z &  \ar@{-}[u]  \ar@{-}[d]  \ar@{-}[l]  \ar@{-}[r] & Z
\\
 & Z
}
\end{array},
&&
\begin{array}{c}
\xymatrix@!0{%
& Z & Z
\\
 &  \ar@{-}[u]  \ar@{-}[d]  \ar@{-}[ur] \ar@{-}[dl]  & 
\\
Z & Z
}
\end{array},
&&
\begin{array}{c}
\xymatrix@!0{%
&  & Z
\\
Z & \ar@{-}[l]  \ar@{-}[r] \ar@{-}[ur] \ar@{-}[dl]  & Z
\\
Z & 
}
\end{array},
&&
\begin{array}{c}
\xymatrix@!0{%
&& & X \ar@{-}[l]\ar@{-}[r] & 
\\
& X \ar@{-}[dl] \ar@{-}[ur]  & X \ar@{.}[u]  & X \ar@{-}[ur] \ar@{-}[dl] & X \ar@{-}[u]\ar@{-}[d]
\\
  & X  \ar@{-}[l]\ar@{-}[r] & & X \ar@{.}[r]  &
\\
 X \ar@{-}[u]\ar@{-}[d] & X \ar@{.}[dl] & X \ar@{-}[u]\ar@{-}[d] & X \ar@{-}[ur] \ar@{-}[dl] 
\\
  & X  \ar@{-}[l]\ar@{-}[r] & 
}
\end{array}. 
\end{align}

\subsection{Fractal spin liquids}

In this subsection we describe coupled 2D (and 1D) cluster state constructions of FSL models. 
We provide several examples: the Sierpinski type-I FSL, a type-II FSL, and cubic code written in FSL form. 
These are followed by a general construction for all FSL models. 

First we construct the Sierpinski FSL from coupled 2D fractal cluster states and 1D cluster states. The starting point is layers of fractal cluster states on $xy$ planes of the cubic lattice, and 1D cluster states stacked along the $\hat{z}$ axis of the cubic lattice. This involves one qubit per $\hat{z}$ edge, and $xy$ plaquette, and two qubits per vertex. The stabilizer generators are 
\begin{align}
\begin{array}{c}
\xymatrix@!0{%
  IZ \ar@{-}[dd]  \ar@{-}[rr]  & & IZ
\\
 & X & 
\\
  IZ & &  \ar@{-}[ll]  \ar@{-}[uu] 
}
\end{array},
&&
\begin{array}{c}
\xymatrix@!0{%
  \ar@{-}[dd]  \ar@{-}[rr]   &   &  \ar@{-}[dd]  \ar@{-}[rr]   &&
\\
   &  &  & Z &
\\
   \ar@{-}[dd]  \ar@{-}[rr]  & & IX  \ar@{-}[dd]  \ar@{-}[rr]  & & \ar@{-}[dd]  \ar@{-}[uu]
\\
  & Z &  & Z &
\\
  &  &   \ar@{-}[ll]  \ar@{-}[rr]  & & 
}
\end{array},
&&
\begin{array}{c}
\xymatrix@!0{%
& & Z 
\\
& XI \ar@{-}[u]  \ar@{-}[d]  \ar@{-}[l]  \ar@{-}[r] \ar@{-}[ur] \ar@{-}[dl]   & 
\\
Z &&
}
\end{array},
&&
\begin{array}{c}
\xymatrix@!0{%
& &  ZI
\\
&  X  \ar@{-}[ur]  \ar@{-}[dl]   &
\\
 ZI &
}
\end{array}. 
\end{align}
We group the $\hat{z}$ edge spins and $xy$ plaquette spins onto the corner of minimal $x,y,z,$ coordinate and add a strong $-(XX+ZZ)$ coupling to these sites. In the limit of infintie coupling we find stabilizer generators 
\begin{align}
\begin{array}{c}
\drawgenerator{}{}{ZI}{}{}{ZZ}{IZ}{IZ}
\end{array},
&&
\begin{array}{c}
\drawgenerator{}{XX}{XI}{XI}{}{}{IX}{}
\end{array}.
\end{align}

We can, in fact, further decompose the 2D fractal cluster states into coupled 1D cluster states with stabilizer generators 
\begin{align}
\begin{array}{c}
\xymatrix@!0{%
  IZI \ar@{-}[dd]  \ar@{-}[rr]  & & IZI
\\
 & XI & 
\\
   & &  \ar@{-}[ll]  \ar@{-}[uu] 
}
\end{array},
&&
\begin{array}{c}
\xymatrix@!0{%
   \ar@{-}[dd]  \ar@{-}[rr]  & & IXI  \ar@{-}[dd]  \ar@{-}[rr]  & & \ar@{-}[dd]
\\
  & ZI &  & ZI &
\\
  &  &   \ar@{-}[ll]  \ar@{-}[rr]  & & 
}
\end{array}
&&
\begin{array}{c}
\xymatrix@!0{%
   \ar@{-}[dd]  \ar@{-}[rr]  & & 
\\
 & IX & 
\\
  IIZ & &  \ar@{-}[ll]  \ar@{-}[uu] 
}
\end{array},
&&
\begin{array}{c}
\xymatrix@!0{%
   \ar@{-}[dd]  \ar@{-}[rr]  & & 
\\
 & IZ & 
\\
  IIX & &  \ar@{-}[ll]  \ar@{-}[uu] 
}
\end{array},
\end{align}
which return the 2D fractal cluster states upon adding a strong $-IZZ$ coupling to the vertices and $-ZZ$ to the faces.

Next we consider a type-II $\mathbb{Z}_2$ FSL. We begin with fractal cluster states stacked in the $xy$ and $xz$ planes of a cubic lattice, with one qubit per $xy$ and $xz$ plaquette, and two qubits per vertex. The stabilizer generators are 
\begin{align}
\begin{array}{c}
\xymatrix@!0{%
  IZ \ar@{-}[dd]  \ar@{-}[rr]  & & IZ \ar@{-}[rr] & &  && IZ \ar@{-}[ll]  \ar@{-}[dd]
\\
 & X & & & && 
\\
  IZ & &  \ar@{-}[ll]  \ar@{-}[uu]  & &  \ar@{-}[ll]  \ar@{-}[uu]  \ar@{-}[rr]  &&
}
\end{array},
&&
\begin{array}{c}
\xymatrix@!0{%
 & & & &   &   &  \ar@{-}[dd]  \ar@{-}[rr]   &&
\\
&&   & & &  &  & Z &
\\
\ar@{-}[dd]  \ar@{-}[rr]   && \ar@{-}[rr] \ar@{-}[dd]  & &  \ar@{-}[dd]  \ar@{-}[rr]  & & IX  \ar@{-}[dd]  \ar@{-}[rr]  & & \ar@{-}[dd]  \ar@{-}[uu]
\\
& Z &  & & & Z &  & Z &
\\
\ar@{-}[rr]   &&   \ar@{-}[rr] & & &  &   \ar@{-}[ll]  \ar@{-}[rr]  & & 
}
\end{array},
\\
\begin{array}{c}
\xymatrix@!0{%
&& ZI \ar@{-}[ddll] \ar@{-}[rr] &  & && \ar@{-}[ddll] \ar@{-}[ll]
\\
& & X & &&
\\
 ZI &  & ZI  \ar@{-}[ll] \ar@{-}[rr] \ar@{-}[uurr] && ZI
}
\end{array},
&&
\begin{array}{c}
\xymatrix@!0{%
& & \ar@{-}[rr] && \ar@{-}[rr] \ar@{-}[ddll] &  & & & \ar@{-}[ll] \ar@{-}[ddll]
\\
& & Z & & Z & & Z & 
\\
\ar@{-}[uurr] \ar@{-}[rr] && & & XI \ar@{-}[ll] \ar@{-}[rr] \ar@{-}[uurr] \ar@{-}[ddll] &  & 
\\
&   & & & Z & 
\\
 & & & & \ar@{-}[ll] \ar@{-}[uurr]
}
\end{array}.
\end{align}
Grouping spins on $xz$ and $xy$ plaquettes to their minimal corner and adding a strong $-(XX+ZZ)$ coupling to these sites, we find the following stabilizer generators in the limit of infinite coupling 
\begin{align}
\begin{array}{c}
\xymatrix@!0{%
& IZ \ar@{-}[ld]\ar@{.}[dd] \ar@{-}[rr] & &  IZ \ar@{-}[ld] & &  \ar@{-}[ld]  \ar@{-}[ll]  \ar@{.}[dd] & & IZ \ar@{-}[ll]  \ar@{-}[dd] \ar@{-}[ld]  \\%
 \ar@{-}[rr] \ar@{-}[dd] &  &  \ar@{-}[dd]  \ar@{-}[rr] &  & &  & \ar@{-}[ll]  \ar@{-}[dd]  &      \\%
 & ZZ  \ar@{.}[ld] &  &  \ar@{.}[uu] \ar@{.}[ll]  \ar@{.}[rr]  & & & & \ar@{-}[dl] \ar@{.}[ll]   \\%
ZI \ar@{-}[rr] &  & ZI  \ar@{.}[ru]  & &  ZI \ar@{-}[ll]  \ar@{-}[uu]  \ar@{.}[ur] \ar@{-}[rr]   & &              %
}%
\end{array},
&&
\begin{array}{c}
\xymatrix@!0{%
& & & IX \ar@{-}[ld]\ar@{.}[dd] \ar@{-}[ll] \ar@{-}[rr] & & IX \ar@{-}[ld] & & IX \ar@{-}[ld]  \ar@{-}[ll]  \ar@{-}[dd]  \\%
\ar@{-}[dd] \ar@{-}[ur]  \ar@{-}[rr]  & & \ar@{-}[rr] \ar@{-}[dd] &  &  \ar@{-}[dd]  \ar@{-}[rr] &  & XX &        \\%
& \ar@{.}[ld] \ar@{.}[uu] \ar@{.}[rr]  & &  \ar@{.}[ld] &  &  \ar@{.}[uu] \ar@{.}[ll]  \ar@{.}[rr]  & &    \\%
XI & &  \ar@{-}[rr] \ar@{-}[ll] & & XI \ar@{.}[ru]  & &  XI \ar@{-}[ll]  \ar@{-}[uu]  \ar@{-}[ur]                    %
}%
\end{array}.
\end{align}
The 2D fractal cluster states can be further broken up into coupled 1D cluster states with stabilizer generators 
\begin{align}
\begin{array}{c}
\xymatrix@!0{%
  IIIZ \ar@{-}[dd]  \ar@{-}[rr]  & &  
\\
 & IX & 
\\
  IIIZ & &  \ar@{-}[ll]  \ar@{-}[uu]  
}
\end{array},
&&
\begin{array}{c}
\xymatrix@!0{%
\ar@{-}[dd]  \ar@{-}[rr]   &&
\\
 & IZ &
\\
IIIX  \ar@{-}[dd]  \ar@{-}[rr]  & & \ar@{-}[dd]  \ar@{-}[uu]
\\
  & IZ &
\\
  \ar@{-}[rr]  & & 
}
\end{array},
\\
\begin{array}{c}
\xymatrix@!0{%
   \ar@{-}[dd]  \ar@{-}[rr]  & & IIZI \ar@{-}[rr] & &  && IIZI \ar@{-}[ll]  \ar@{-}[dd]
\\
 & XI & & & && 
\\
   & &  \ar@{-}[ll]  \ar@{-}[uu]  & &  \ar@{-}[ll]  \ar@{-}[uu]  \ar@{-}[rr]  &&
}
\end{array},
&&
\begin{array}{c}
\xymatrix@!0{%
\ar@{-}[dd]  \ar@{-}[rr]   && \ar@{-}[rr] \ar@{-}[dd]  & &  \ar@{-}[dd]  \ar@{-}[rr]  & & IIXI  \ar@{-}[dd]  
\\
& ZI &  & & & ZI &
\\
\ar@{-}[rr]   &&   \ar@{-}[rr] & & &  &   \ar@{-}[ll] 
}
\end{array},
\end{align}
\begin{align}
\begin{array}{c}
\xymatrix@!0{%
&&  \ar@{-}[ddll] \ar@{-}[rr] &  & && \ar@{-}[ddll] \ar@{-}[ll]
\\
& & XI & &&
\\
  &  & ZIII  \ar@{-}[ll] \ar@{-}[rr] \ar@{-}[uurr] && ZIII
}
\end{array},
&&
\begin{array}{c}
\xymatrix@!0{%
& & \ar@{-}[rr] && \ar@{-}[rr] \ar@{-}[ddll] &  & 
\\
& & ZI & & ZI & 
\\
\ar@{-}[uurr] \ar@{-}[rr] && & & XIII \ar@{-}[ll]  \ar@{-}[uurr]
}
\end{array},
\\
\begin{array}{c}
\xymatrix@!0{%
&& IZII \ar@{-}[ddll] \ar@{-}[rr] &  &
\\
& & IX & 
\\
 IZII &  & \ar@{-}[ll] \ar@{-}[uurr]
}
\end{array},
&&
\begin{array}{c}
\xymatrix@!0{%
 &&  &  & & & \ar@{-}[ll] \ar@{-}[ddll]
\\
& &  & & IZ & 
\\
 & & IXII \ar@{-}[rr] \ar@{-}[uurr] \ar@{-}[ddll] &  & 
\\
 & & IZ & 
\\
 & & \ar@{-}[ll] \ar@{-}[uurr]
}
\end{array}.
\end{align}
In the limit of a strong $-ZZ$ coupling on the faces, and $-(IIZZ + ZZII)$ on the vertices we recover the original cluster states.

Next we consider the cubic code written in FSL form~\cite{yoshida2013exotic}. It is constructed from coupled fractal cluster states stacked along the $xy$ and $xz$ planes of the cubic lattice, with one qubit per $xy$ and $xz$ plaquette, and two qubits per vertex. The stabilizer generators are 
\begin{align}
\begin{array}{c}
\xymatrix@!0{%
  IZ \ar@{-}[dd]  \ar@{-}[rr]  & & IZ \ar@{-}[rr] \ar@{-}[dd] & & IZ \ar@{-}[dd] 
\\
 & & & & 
\\
  IZ \ar@{-}[dd]  \ar@{-}[rr]  & & IZ \ar@{-}[rr] & & 
\\
 & X & & & 
\\
  IZ & &  \ar@{-}[ll]  \ar@{-}[uu]  & &  \ar@{-}[ll]  \ar@{-}[uu] 
}
\end{array},
&&
\begin{array}{c}
\xymatrix@!0{%
& &     &   &  \ar@{-}[dd]  \ar@{-}[rr]   &&
\\
   & & &  &  & Z &
\\
   & &  \ar@{-}[dd]  \ar@{-}[rr]  & & IX  \ar@{-}[dd]  \ar@{-}[rr]  & & \ar@{-}[dd]  \ar@{-}[uu]
\\
  &  & & Z &  & Z &
\\
  \ar@{-}[rr] & & &  &   \ar@{-}[ll]  \ar@{-}[rr]  & & 
\\
  & Z & & Z &  & Z &
\\
  \ar@{-}[rr] \ar@{-}[uu] & & \ar@{-}[uu] &  &  \ar@{-}[uu] \ar@{-}[ll]  \ar@{-}[rr]  & & \ar@{-}[uu]
}
\end{array},
\\
\begin{array}{c}
\xymatrix@!0{%
&& ZI \ar@{-}[ddll] \ar@{-}[rr] &  & && \ar@{-}[ddll] \ar@{-}[ll]
\\
& & X & &&
\\
 ZI &  & ZI  \ar@{-}[ll] \ar@{-}[rr] \ar@{-}[uurr] && ZI
}
\end{array},
&&
\begin{array}{c}
\xymatrix@!0{%
& & \ar@{-}[rr] && \ar@{-}[rr] \ar@{-}[ddll] &  & & & \ar@{-}[ll] \ar@{-}[ddll]
\\
& & Z & & Z & & Z & 
\\
\ar@{-}[uurr] \ar@{-}[rr] && & & XI \ar@{-}[ll] \ar@{-}[rr] \ar@{-}[uurr] \ar@{-}[ddll] &  & 
\\
&   & & & Z & 
\\
 & & & & \ar@{-}[ll] \ar@{-}[uurr]
}
\end{array}.
\end{align}
Grouping spins on $xz$ and $xy$ plaquettes to their minimal corner and adding a strong $-(XX+ZZ)$ coupling to these sites, in the limit of infinite coupling strength we find the stabilizer generators of cubic code in FSL form 
\begin{align}
\begin{array}{c}
\xymatrix@!0{%
& IZ \ar@{-}[ld]\ar@{.}[dd] \ar@{-}[rr] & &  IZ \ar@{-}[ld] & & IZ \ar@{-}[ld]  \ar@{-}[ll]  \ar@{-}[dd]  \\%
 \ar@{-}[rr] \ar@{-}[dd] &  &  \ar@{-}[dd]  \ar@{-}[rr] &  &  \ar@{-}[dd]  &  \\%
& IZ \ar@{.}[ld]\ar@{.}[dd] \ar@{.}[rr] & &  IZ \ar@{.}[ld]  \ar@{.}[uu] & &  \ar@{-}[ld]  \ar@{.}[ll]  \ar@{-}[dd]  \\%
 \ar@{-}[rr] \ar@{-}[dd] &  &  \ar@{-}[dd]  \ar@{-}[rr] &  & &       \\%
 & ZZ  \ar@{.}[ld] &  &  \ar@{.}[uu] \ar@{.}[ll]  \ar@{.}[rr]  & &   \\%
ZI \ar@{-}[rr] &  & ZI  \ar@{.}[ru]  & &  ZI \ar@{-}[ll]  \ar@{-}[uu]  \ar@{-}[ur]             %
}%
\end{array},
&&
\begin{array}{c}
\xymatrix@!0{%
 & IX \ar@{-}[ld]\ar@{.}[dd] \ar@{-}[rr] & & IX \ar@{-}[ld] & & IX \ar@{-}[ld]  \ar@{-}[ll]  \ar@{-}[dd]  \\%
 \ar@{-}[rr] \ar@{-}[dd] &  &  \ar@{-}[dd]  \ar@{-}[rr] &  & XX &        \\%
 &  \ar@{.}[dd]  \ar@{.}[ld] &  &  \ar@{.}[uu] \ar@{.}[ll]  \ar@{.}[rr]  & &    \\%
  \ar@{-}[rr]  & & XI \ar@{.}[ru]  & &  XI \ar@{-}[ll]  \ar@{-}[uu]  \ar@{-}[ur]             \\%
 &  \ar@{.}[ld] &  &  \ar@{.}[uu] \ar@{.}[ll]  \ar@{.}[rr]  & & \ar@{-}[uu]   \\%
XI  \ar@{-}[uu] \ar@{-}[rr]  & & XI \ar@{-}[uu] \ar@{.}[ru]  & &  XI \ar@{-}[ll]  \ar@{-}[uu]  \ar@{-}[ur]              %
}%
\end{array}.
\end{align}

\subsubsection{General fractal spin liquids}


Finally, we describe a general coupled cluster layer construction for any FSL. This was depicted schematically in Fig.~\ref{fig:FSL}, where $xy$ fractal cluster layers and $xz$ fractal cluster layers are strongly coupled to produce the FSL model. 
The Hamiltonian, or stabilizer map, for any FSL can be written using the polynomial stabilizer formalism~\cite{Haah2013,haah2013commuting,yoshida2013exotic}
\begin{align}
	\begin{pmatrix}
	0 & G(\bar{x},\bar{z})  \\
	0 &  F(\bar{x},\bar{y}) \\
	F(x,y) & 0 \\
	G(x,z) & 0 
	\end{pmatrix} ,
	\label{eq:polynomialfsl}
\end{align}
where $ F(x,y)  = 1+f(x)y$, $G(x,z)=1+g(x)z$ for a first order FSL. For more general higher order FSL models we have 
\begin{align}
F(x,y)  = 1+\sum_{i\geq 1} f_i(x) y^i \, , && G(x,y)  = 1+\sum_{i\geq 1} g_i(x) z^i \, .
\end{align}

We consider a stack of fractal cluster states along the $\hat{y}$ and $\hat{z}$ axes
\begin{align}
    \begin{pmatrix}
    1 & 0 & 0 & 0 \\
    0 & 1 & 0 & 0 \\
    0 & 0 & 1 & 0 \\
    0 & 0 & 0 & 1 \\
    0 & F(\bar{x},\bar{y}) & 0 & 0 \\
    F(x,y) & 0 & 0 & 0 \\
    0 & 0 & 0 & G(\bar{x},\bar{z}) \\
    0 & 0 & G(x,z) & 0 
    \end{pmatrix} . 
    \label{eq:stabgen} 
\end{align}
We point out that the first and second qubits on every site are decoupled from the third and fourth and the Hamiltonian terms acting on each pair of qubits involve only two variables. Hence the above Hamiltonian corresponds to a stack of decoupled layers. 

Next we introduce strong on-site field couplings described by
\begin{align}
    \begin{pmatrix}
    1 & 0 \\
    0 & 0 \\
    1 & 0 \\
    0 & 0 \\
    0 & 1 \\
    0 & 0 \\
    0 & 1 \\
    0 & 0 
    \end{pmatrix} . 
\end{align}
To find the leading order Hamiltonian terms in perturbation theory we construct a minimal set of stabilizer generators that commute with the strong fields via linear combinations of columns from Eq.~\eqref{eq:stabgen} 
\begin{align}
    \begin{pmatrix}
    1 & 0   \\
    0 & G(\bar{x},\bar{z}) \\
    1 & 0  \\
    0 & F(\bar{x},\bar{y})  \\
    0 & F(\bar{x},\bar{y}) G(\bar{x},\bar{z})  \\
    F(x,y) & 0  \\
    0 & F(\bar{x},\bar{y}) G(\bar{x},\bar{z})  \\
    G(x,z) & 0 
    \end{pmatrix} . 
\end{align}
Removing the first and third qubits from every site that are fixed out by the strong fields corresponds to dropping the odd numbered rows from the above matrix. This recovers the general FSL model in Eq.~\eqref{eq:polynomialfsl}. 

The layered construction described here generalizes straightforwardly to FSL models involving more than two qubits per site. In fact it generalizes to any other model where the qubits can be partitioned into two disjoint subsets such that the $Z$-type stabilizer terms act on the first subset via functions of only $(x,y)$ and on the second subset via functions of only $(x,z)$. 
For an example of this generalization see cubic code B below.

\subsection{Cubic code B}

In this subsection we describe a construction of cubic code B from coupled 2D fractal cluster states stacked in $xy$ and $xz$ planes of the cubic lattice. The decoupled system consists of two qubits per $xy$ and $xz$ plaquette of the cubic lattice, and four qubits per vertex. The stabilizer generators are 
\begin{align}
\begin{array}{c}
\xymatrix@!0{%
  IIZZ  \ar@{-}[dd]  \ar@{-}[rr]  & & 
\\
 & XI & 
\\
  IIIZ & & IIZI \ar@{-}[ll]  \ar@{-}[uu] 
}
\end{array},
&&
\begin{array}{c}
\xymatrix@!0{%
  IIZI  \ar@{-}[dd]  \ar@{-}[rr]  & & 
\\
 & IX & 
\\
  IIZZ & & IIIZ \ar@{-}[ll]  \ar@{-}[uu] 
}
\end{array},
&&
\begin{array}{c}
\xymatrix@!0{%
  \ar@{-}[dd]  \ar@{-}[rr]   &   &  \ar@{-}[dd]  \ar@{-}[rr]   &&
\\
   & ZI &  & IZ &
\\
   \ar@{-}[dd]  \ar@{-}[rr]  & & IIXI  \ar@{-}[dd]  \ar@{-}[rr]  & & \ar@{-}[dd]  \ar@{-}[uu]
\\
  &  &  & ZZ &
\\
  &  &   \ar@{-}[ll]  \ar@{-}[rr]  & & 
}
\end{array},
&&
\begin{array}{c}
\xymatrix@!0{%
  \ar@{-}[dd]  \ar@{-}[rr]   &   &  \ar@{-}[dd]  \ar@{-}[rr]   &&
\\ 
   & IZ &  & ZZ &
\\
   \ar@{-}[dd]  \ar@{-}[rr]  & & IIIX  \ar@{-}[dd]  \ar@{-}[rr]  & & \ar@{-}[dd]  \ar@{-}[uu]
\\
  &  &  & ZI &
\\
  &  &   \ar@{-}[ll]  \ar@{-}[rr]  & & 
}
\end{array},
\end{align}
\begin{align}
\begin{array}{c}
\xymatrix@!0{%
&& IZII \ar@{-}[ddll] \ar@{-}[rr] &  & ZZII
\\
& & XI &
\\
 ZIII &  &  \ar@{-}[ll] \ar@{-}[uurr]
}
\end{array},
&&
\begin{array}{c}
\xymatrix@!0{%
&& ZZII \ar@{-}[ddll] \ar@{-}[rr] &  & ZIII
\\
& & IX &
\\
 IZII &  &  \ar@{-}[ll] \ar@{-}[uurr]
}
\end{array},
\\
\begin{array}{c}
\xymatrix@!0{%
& & && \ar@{-}[rr] \ar@{-}[ddll] &  & & & \ar@{-}[ll] \ar@{-}[ddll]
\\
& & & &  & & ZI & 
\\
&& & & XIII \ar@{-}[ll] \ar@{-}[rr] \ar@{-}[uurr] \ar@{-}[ddll] &  & 
\\
&   & ZZ & & IZ & 
\\
\ar@{-}[uurr] \ar@{-}[rr]  & & & & \ar@{-}[ll] \ar@{-}[uurr]
}
\end{array},
&&
\begin{array}{c}
\xymatrix@!0{%
& & && \ar@{-}[rr] \ar@{-}[ddll] &  & & & \ar@{-}[ll] \ar@{-}[ddll]
\\
& & & &  & & IZ & 
\\
&& & & IXII \ar@{-}[ll] \ar@{-}[rr] \ar@{-}[uurr] \ar@{-}[ddll] &  & 
\\
&   & ZI & & ZZ & 
\\
\ar@{-}[uurr] \ar@{-}[rr]  & & & & \ar@{-}[ll] \ar@{-}[uurr]
}
\end{array}.
\end{align}
Grouping the spins on $xz$ and $xy$ faces onto the corner of minimal $x,y,z,$ coordinate and adding a strong $-(XIXI+IXIX+ZIZI+IZIZ)$ coupling to these sites we find the stabilizer generators for cubic code B (with a spatial rotation compared to Ref.~\onlinecite{haah2014bifurcation})
\begin{align}
\begin{array}{c}
\drawgenerator{}{}{ZIII}{}{ZZZI}{IZIZ}{}{IIZZ}
\end{array},
&&
\begin{array}{c}
\drawgenerator{}{}{IZII}{}{ZIIZ}{ZZZZ}{}{IIZI}
\end{array},
&&
\begin{array}{c}
\drawgenerator{XIXX}{IXIX}{}{XXII}{}{}{IIXI}{}
\end{array},
&&
\begin{array}{c}
\drawgenerator{IXXI}{XXXX}{}{XIII}{}{}{IIIX}{}
\end{array}.
\end{align}

Similar to the above examples, the 2D fractal cluster states can be further decomposed into coupled 1D cluster states.

\end{document}